\documentclass[twocolumn]{aastex701}

\usepackage{amsmath}
\usepackage{amsfonts}
\usepackage{amssymb}
\usepackage{array}
\usepackage{indentfirst}
\usepackage{comment}
\usepackage{mathtools}
\usepackage{graphicx}
\usepackage{booktabs}
\usepackage{placeins}
\begin{document}

\title{The Incidence of Large Ionized Bubbles at Redshift 13}

\author[0009-0008-9919-6619]{Peter Ziwei Hu}
\affiliation{The William H. Miller III Department of Physics \& Astronomy, Johns Hopkins University, Baltimore, MD, USA}
\email[]{zhu20@jhu.edu}

\author[0000-0001-9935-6047]{Massimo Stiavelli}
\affiliation{Space Telescope Science Institute, 3700 San Martin Drive, Baltimore, MD 21218, USA}
\affiliation{The William H. Miller III Department of Physics \& Astronomy, Johns Hopkins University, Baltimore, MD, USA}
\email[]{mstiavel@stsci.edu}

\author[0000-0002-5222-5717]{Colin Norman}
\affiliation{The William H. Miller III Department of Physics \& Astronomy, Johns Hopkins University, Baltimore, MD, USA}
\affiliation{Space Telescope Science Institute, 3700 San Martin Drive, Baltimore, MD 21218, USA}
\email[]{cnorman3@jhu.edu}

\correspondingauthor{Peter Ziwei Hu}
\email[show]{zhu20@jhu.edu}

\begin{abstract}
Ionized bubbles around the first galaxies connect early galaxy growth, ionizing photon escape, intergalactic-medium topology, Ly$\alpha$ visibility, 21~cm structure, and reionization timing. We model the incidence of galaxy-driven ionized bubbles at $z\approx 13$ using \textit{JWST} UV luminosity functions, taking the Ly$\alpha$ source reported by \citet{Witstok2025Nature} as a benchmark for the relevant scale. We quantify regions with comoving radius $R\ge 2.5$~cMpc through their sky surface density $\Sigma_{\ge 2.5}$. For our continuously active fiducial case, using the \citet{Donnan2024PRIMER} UVLF with $f_{\mathrm{esc}}=0.2$, $\log\xi_{\mathrm{ion}}=25.5$, $f_{\mathrm{active}}=1$, and $C=3$, we find $\Sigma_{\ge 2.5}\simeq 8.10\times 10^{-3}$ arcmin$^{-2}$ per $\Delta z=1$. For $f_{\rm active}=0.2$ or $f_{\rm time}=0.2$ with 10~Myr mean on-times, the incidence falls by factors of 4.1 and 2.9, respectively; longer correlated episodes strengthen the tail. Merging overlapping spheres raises the fiducial incidence by $9.5\%$. The simulations show that Witstok-sized regions can arise in UVLF-calibrated galaxy-driven models. This is a population-level statement, and the specific Witstok source may still require unusual effective ionizing efficiency, recent fading or burstiness, or a non-stellar ionizing contribution.
\end{abstract}


\section{Introduction}\label{sec:intro}

Ionized bubbles around early galaxies are the spatial units through which reionization proceeds. Their sizes and topology connect the galaxy population to the escape of ionizing photons, the timing of reionization, Ly$\alpha$ transmission, 21~cm structure, and the integrated CMB optical depth \citep{Furlanetto2004,McQuinn2007,Zahn2007,FurlanettoOh2016,Giri2018,Planck2018,Pagano2020}. The abundance of large bubbles therefore informs both galaxy formation and cosmology, as well as the interpretation of individual unusual sources.

The bubble distribution depends on both the sources (how many galaxies form, how bright they are, and how efficiently they leak ionizing photons) and the sinks (recombinations in dense gas). Large-box radiative-transfer calculations explore how source populations and gas structure shape reionization topology \citep{Ocvirk2020,Kannan2022THESAN,Garaldi2024}. Here we isolate the sensitivity of the high-radius tail to the abundance of luminous galaxies and their effective ionizing emissivity.

\textit{JWST} has reshaped the source side of this problem. Deep imaging and spectroscopy now constrain the bright end of the UV luminosity function out to $z\gtrsim 10$--14, and many analyses report a higher abundance of luminous galaxies than pre-\textit{JWST} extrapolations predicted \citep{Finkelstein2023CEERS,Harikane2025JWST,Donnan2024PRIMER}. Spectroscopic measurements of the ionizing photon production efficiency $\xi_{\rm ion}$ \citep{Simmonds2024,Atek2024} and indirect constraints on $f_{\rm esc}$ \citep{Mascia2023} motivate the range of effective $f_{\rm esc}\xi_{\rm ion}$ products explored here.

Ly$\alpha$ detections in GN-z11 and several $z>8$ galaxies provide context for the range of source histories and ionized environments inferred at early epochs \citep{Bunker2024GNz11,Witstok2025JADESLAEs,Hu2026C3PO}. JADES-GS-z13-1-LA is distinguished by substantial Ly$\alpha$ emission (rest-frame equivalent width $>40$~\AA) at $z\approx13$ \citep{Witstok2025Nature}. We use it as a benchmark for the ionized-region scale, without assuming that the source itself is typical, and compare it with other emitters.

The Ly$\alpha$ line is resonantly absorbed by neutral hydrogen and is hard to see through a neutral intergalactic medium; an ionized environment can increase transmission by allowing the line to redshift before encountering the predominantly neutral IGM. The inferred distance depends on the intrinsic line profile, velocity offset, and intervening neutral gas \citep{Witstok2025Nature,Bunker2024GNz11,Mason2025}. The question we ask is whether baseline galaxy-driven models, tied to \textit{JWST} UVLFs, already predict a non-negligible abundance of such large local ionized environments.

An important definitional distinction is that Ly$\alpha$ transmission is sensitive to the distance to the nearest neutral gas along the line of sight, whereas ``bubble size'' in simulations can mean anything from a single-source Str\"omgren sphere to the extent of a connected ionized component \citep{Lin2016,Giri2018}. Bubble size distributions conditioned on galaxy luminosity and environment further distinguish the ionized regions around selected galaxies from population-wide counts \citep{Lu2024BSD}. We use the comoving radius $R$ of each non-merging source-group sphere as a proxy for the line-of-sight transmission distance. Distinct sphere objects may overlap but are not merged in the fiducial calculation, so the connected ionized scale around clustered sources can be larger than what we report, although the mapping to any particular Ly$\alpha$ sight line remains geometry-dependent (see \S\ref{sec:discussion_mergers} for the merger calculation). We focus on the incidence of ``Witstok-sized'' environments at $z\approx 13$, defined as bubbles with comoving radius $R\ge 2.5$~cMpc and quantified by the sky surface density $\Sigma_{\ge 2.5}$ (arcmin$^{-2}$ per $\Delta z=1$). This numerical threshold comes from the modeling of JADES-GS-z13-1-LA in \citet{Witstok2025Nature}, where the object is inferred to have a local ionized bubble of approximately $R_{\rm ion}\simeq0.18$ physical Mpc, corresponding to approximately $2.5$~cMpc at $z\simeq13$. The deeper OASIS observations give a model-derived radius of $0.167^{+0.070}_{-0.062}$ physical Mpc at $z=13.07$, consistent with the original 2.5~cMpc benchmark retained here \citep{Witstok2026OASIS}.

Section~\ref{sec:model} describes the simulation framework. Section~\ref{sec:results} presents the fiducial incidence, parameter controls, global ionization context, and bubble size distribution evolution. Section~\ref{sec:discussion} connects the population-level result to the Witstok object, source-activity sensitivity, and merger/percolation effects, and Section~\ref{sec:conclusions} summarizes the implications.

We adopt the Planck 2018 cosmology \citep{Planck2018} for the simulations and analytic calculations. Distances are comoving unless otherwise stated.

\section{Model Description}\label{sec:model}

We use a UVLF-calibrated Monte Carlo model to follow the growth of ionized regions from $z=25$. The fiducial calculation evolves non-merging source-group spheres and includes (1) galaxy population synthesis from the observed UV luminosity function and (2) bubble expansion with recombinations and source-activity effects. Spatial ``painting'' maps the ionized spheres onto a grid to estimate the ionized fraction in each cell. This optional measurement and merger tracking under periodic boundaries do not enter the fiducial incidence calculation. Table~\ref{tab:params} summarizes the physical and numerical parameters.

\begin{deluxetable*}{lllll}
\tabletypesize{\scriptsize}
\setlength{\tabcolsep}{3pt}
\tablecaption{Key model parameters and fiducial values\label{tab:params}}
\tablehead{
\colhead{Symbol / Name} & \colhead{Meaning} & \colhead{Fiducial} & \colhead{Range (tests)} & \colhead{Notes}
}
\startdata
$L_{\rm box}$, $N$  &  Box size, grid cells  &  $50$ cMpc, $128^3$  & $8\times250$ cMpc boxes & --- \\
$z_{\rm start},z_{\rm end}$; $\Delta t$  &  Redshift span; time step  &  $25,5$; $5$~Myr  & $z_{\rm start}=15.5,20$ & Rare-tail runs end at $z=12$ \\
$M_{\rm lim}$  &  UVLF faint-end limit  &  $-13$  & $[-14.5,-12.0]$ & Steps of $0.5$ mag \\
$\{\phi^\star,M^\star,\alpha,\beta\}(z)$  &  DPL UVLF params  &  D24  & B21 & Schechter comparison \\
$f_{\rm esc}$  &  Escape fraction  &  $0.2$  & $[0.1,0.4]$ & Also C22+C24, A17, N20+S15 \\
$\xi_{\rm ion}$  &  Ionizing efficiency  &  $10^{25.5}$ Hz erg$^{-1}$  & $\pm1.0$ dex & Also S24a, S24b \\
$f_{\rm active}$  &  Population active fraction  &  $1.0$  & $[0.1,1.0]$ & Instantaneous occupancy \\
$f_{\rm time}$  &  Stationary on fraction  &  ---  & $\{0.1,0.2,0.4,0.5,0.6,0.8\}$ & $\langle t_{\rm on}\rangle=10$ Myr \\
$\langle t_{\rm on}\rangle$  &  Mean active-episode duration  &  ---  & $\{2.5,10,40\}$ Myr & $f_{\rm time}=0.2$ \\
$C$  &  Clumping factor  &  $3$  & $\{1,20,\mathrm{MD14}\}$ & Constant or redshift dependent \\
\enddata
\tablecomments{D24 = \citet{Donnan2024PRIMER}; B21 = \citet{Bouwens2021}; MD14 = \citet{MadauDickinson2014}. Luminosity-dependent tests: C22+C24 = \citet{Chisholm2022} with \citet{Cullen2024}; A17 = \citet{Anderson2017}; N20+S15 = \citet{Naidu2020} with \citet{Shibuya2015}; S24a = \citet{Simmonds2024MassComplete}; S24b = \citet{Simmonds2024}. Activity models are tested separately for D24 and B21; $f=1$ is the continuous limit. T($t_{\rm on}$) denotes time duty in the results tables. We also compare separate source groups with spherical merging.}
\end{deluxetable*}

We use two sets of simulations for different measurements. Full-grid $L=50$~cMpc, $128^3$ runs extend from $z=25$ to 5 and provide the ionization history, slices, terminal ionized fractions, and the time-resolved merger histories discussed in \S\ref{sec:discussion_mergers}. Main-text rare-tail number densities, surface densities, bubble size distributions, source-activity comparisons, and $z=13$ merger comparisons use eight independent $L=250$~cMpc runs from $z=25$ to 12; their stacked volume is $(500~\mathrm{cMpc})^3$. Both sets use the same physical model and exact snapshot-only outputs at the redshifts entering the incidence calculation.

\subsection{Galaxy population from the UVLF}
Galaxies are seeded from the evolving DPL redshift-prior parameterization of \citet[][Eqs.~3--6]{Donnan2024PRIMER}. The observed sample spans $8.5<z<15.5$, while the adopted redshift prior extends to $z=20$. Predictions above $z=15.5$ therefore extend beyond the observed sample. We continue the DPL branch to $z=25$ and below $z=7$, using the latter continuation only for the full ionization history. Section~\ref{sec:headline_z13} reports the sensitivity to the starting redshift.

The comparison model uses the evolving Schechter form of \citet[][Section~4.2]{Bouwens2021}, extrapolated to $z\approx13$. Both populations span $-30\le M_{\rm UV}\le M_{\rm lim}$.

We update the population every 5~Myr, retaining existing sources and adding galaxies to follow the evolving UVLF. New positions are uniformly random. This omits the enhanced clustering of rare bright galaxies, which can increase connected ionized regions and line-of-sight path lengths.

The rest-frame UV specific luminosity $L_{\mathrm{UV}}$ of each galaxy, in erg~s$^{-1}$~Hz$^{-1}$ and conventionally evaluated near 1500~\AA, is computed from its absolute magnitude, and the ionizing photon production rate is derived as
\begin{equation}
\dot{N}_{\mathrm{ion}} = f_{\mathrm{esc}}\, \xi_{\mathrm{ion}}\, L_{\mathrm{UV}},
\label{eq:ion_rate}
\end{equation}
where $f_{\mathrm{esc}}$ is the escape fraction and $\xi_{\mathrm{ion}}$ is the ionizing photon production efficiency.

Each source retains its assigned $M_{\rm UV}$ and on-state $L_{\rm UV}$ after birth. The activity prescriptions below interrupt emission but do not evolve stellar mass or follow UV brightening and fading through a burst. The resulting incidence is conditional on these effective source histories.

\subsection{Bubble growth and recombination}
The simulation stores each bubble volume and radius in comoving units, with volume evolving according to
\begin{equation}
\begin{aligned}
\frac{\mathrm{d}V}{\mathrm{d}t}
&= \frac{\dot{N}_{\mathrm{ion}}}{\bar{n}_{\mathrm{H},0}}
- \alpha_{\mathrm{B}}\,C\,\bar{n}_{\mathrm{H},0}(1+z)^3 V \\
&\equiv \frac{\dot{N}_{\mathrm{ion}}}{\bar{n}_{\mathrm{H},0}} - \frac{V}{t_{\rm rec}(z)},
\end{aligned}
\label{eq:bubble_growth}
\end{equation}
with $R \equiv (3V/4\pi)^{1/3}$. Here $V$ and $R$ are comoving volume and radius, $\bar{n}_{\mathrm{H},0}$ is the mean comoving hydrogen number density, $n_{\mathrm{H}}(z)=\bar{n}_{\mathrm{H},0}(1+z)^3$ is the corresponding physical hydrogen number density, $\alpha_{\mathrm{B}}$ is the case-B recombination coefficient, and $C$ is the clumping factor. The recombination time is therefore
\begin{equation}
t_{\rm rec}(z) \equiv \left[\alpha_{\mathrm{B}}\,C\,n_{\mathrm{H}}(z)\right]^{-1}
=\left[\alpha_{\mathrm{B}}\,C\,\bar n_{\mathrm{H},0}(1+z)^3\right]^{-1}.
\label{eq:t_rec}
\end{equation}
We use $\alpha_{\mathrm{B}}=2.6\times10^{-13}$~cm$^{3}$~s$^{-1}$ at $T=10^4$~K, a hydrogen mass fraction $X_{\rm H}=0.75$, and a fixed hydrogen-density normalization $\bar n_{\mathrm{H},0}=1.894\times10^{-7}$~cm$^{-3}$, used in both the simulations and analytic calculations. The growth equation treats hydrogen alone; unresolved sinks are absorbed into the effective clumping factor $C$, following standard clumping-factor prescriptions used in reionization modeling \citep{Pawlik2009,Finlator2012,Sobacchi2014}. We also test $C(z)=1+43z^{-1.71}$ from \citet{MadauDickinson2014}, extrapolated below $z=6$.

In the active-fraction model, $f_{\rm active}$ is the target fraction of galaxies that emit at each update; the total population includes both active and inactive galaxies. Every 5~Myr, we independently select active members in each magnitude bin to reproduce the realized UVLF counts. Inactive intervals permit partial recombination. Rounding and retained sources can make the realized occupancy differ from $f_{\rm active}$.

We also test temporal coherence with a time-duty model, in which each galaxy follows a two-state Markov process. The stationary on fraction $f_{\rm time}$ and mean on-time $\langle t_{\rm on}\rangle$ determine the mean off-time,
\begin{equation}
\langle t_{\rm off}\rangle=\langle t_{\rm on}\rangle\frac{1-f_{\rm time}}{f_{\rm time}},
\label{eq:time_duty_relation}
\end{equation}
and both episode durations are exponentially distributed. The larger latent population preserves the expected emitting UVLF up to rounding when no surplus is retained, while individual histories remain correlated across updates. If a UVLF target decreases, retaining previously created galaxies can leave a surplus latent population, so the expected active count can exceed the new target. We run the two activity prescriptions separately.

\subsection{Bubble definition and mergers}
Each newly placed source joins the largest sphere enclosing its birth position; otherwise it starts a zero-volume sphere centered on itself. The fiducial sphere centers remain fixed, and subsequent growth does not reassign existing sources. Each sphere grows through its members' summed emission. The fiducial bubble size distribution and incidence count these spheres separately, including residual volumes during inactive intervals. Fossil volumes describe a remaining hydrogen ionization budget; their Ly$\alpha$ transparency also depends on the residual neutral fraction.

For the ionization history and spatial slices, we estimate the occupied volume on a grid. A separate merger calculation replaces connected sets of overlapping spheres with one sphere whose volume equals the summed parent volumes. Overlap volume is therefore counted in full. These geometric descriptions omit anisotropic radiative transfer and peculiar velocities.

\subsection{Parameter combinations and degeneracies}\label{sec:param_degeneracies}
The cube-root dependence follows directly from photon counting: before recombinations matter, the ionized volume is proportional to the cumulative number of emitted ionizing photons, $V\propto \dot N_{\rm ion}t$, while spherical geometry gives $R\propto V^{1/3}$. The growth law therefore yields two useful comoving-radius scalings: (i) the \emph{photon-counting} regime $R(t)\propto \big[\dot N_{\rm ion}\,t/\bar n_{\rm H,0}\big]^{1/3}$ at $t\ll t_{\rm rec}$, and (ii) the \emph{recombination-limited} asymptote obtained by setting Eq.~\eqref{eq:bubble_growth} to zero,
$R_{\rm S}\propto \big[\dot N_{\rm ion}/(\alpha_{\rm B} C\bar n_{\rm H,0}^{2}(1+z)^3)\big]^{1/3}$, equivalently $R_{\rm S}\propto \big[\dot N_{\rm ion}/(\alpha_{\rm B} C\bar n_{\rm H,0} n_{\rm H}(z))\big]^{1/3}$. Radii therefore depend on $(f_{\rm esc}\,\xi_{\rm ion}\,L_{\rm UV})^{1/3}$. At fixed source luminosity, $f_{\rm esc}$ and $\xi_{\rm ion}$ are exactly degenerate in this model; changing $M_{\rm lim}$ instead changes the population and total UV luminosity density and is not an equivalent object-level rescaling. External priors (e.g., nebular-line $\xi_{\rm ion}$, SEDs, or direct $f_{\rm esc}$ constraints) are required to separate $f_{\rm esc}$ from $\xi_{\rm ion}$. In the simulation, both are effective parameters that absorb uncertain ISM and radiative-transfer physics \citep{Mason2018,Tang2024}.

\section{Results}\label{sec:results}

This section first reports the fiducial $z\approx 13$ incidence of $R\ge 2.5$~cMpc bubbles. It then asks what controls the high-radius tail, places the model in the context of the global ionization history and morphology, and compares the simulated bubble size distribution (BSD) with analytic expectations.

\subsection[Witstok-sized bubbles at z~13]{Witstok-sized bubbles at $z\approx 13$}\label{sec:headline_z13}
We define a ``Witstok-sized'' region as a comoving spherical bubble with $R\ge 2.5$~cMpc at $z\approx 13$ and summarize its incidence using the sky surface density $\Sigma_{\ge 2.5}$. For a threshold $R_0$, the conversion from number density to sky surface density is
\begin{equation}
\Sigma_{\ge R_0}(z)
=
n_{\ge R_0}(z)
\frac{\mathrm{d}V_{\rm com}}{\mathrm{d}\Omega\,\mathrm{d}z}.
\end{equation}

For the finite-window estimate, we compute
\begin{equation}
\Sigma_{\ge R_0}^{\rm 5\,slice}
=
\sum_i n_{\ge R_0}(z_i)
\left.
\frac{\mathrm{d}V_{\rm com}}{\mathrm{d}\Omega\,\mathrm{d}z} \right|_{z_i} \Delta z_i,
\end{equation}
where the sum runs over exact snapshot-only outputs at $z_i=\{12.6,12.8,13.0,13.2,13.4\}$, with integration weights $\Delta z_i=0.2$, to approximate a $\Delta z=1$ window centered on $z\simeq 13$. The main text adopts this five-slice value because it more directly represents the finite redshift interval used in the incidence estimate; the corresponding midpoint-times-$\Delta z$ normalization is reported in Table~\ref{tab:sim_bubbles_full}.

For the continuously active fiducial model (Table~\ref{tab:params}), Table~\ref{tab:sim_bubbles} gives $\Sigma_{\ge 2.5}\simeq 8.10\times 10^{-3}$ at $z\approx 13$. Per unit redshift interval, this corresponds to $N\sim 0.32$ Witstok-sized regions in a JADES-class field ($\sim 40$~arcmin$^2$) and $N\sim 15$ in a COSMOS-Web-scale field ($\sim 0.5$~deg$^2$). This surface density measures the incidence of ionized environments. A quantitative comparison with detection counts would require an emitter-occupation, line-of-sight transmission, and survey-selection model and is beyond the scope of this calculation.

\begin{figure*}[!htb]
    \centering
    \includegraphics[width=0.96\textwidth]{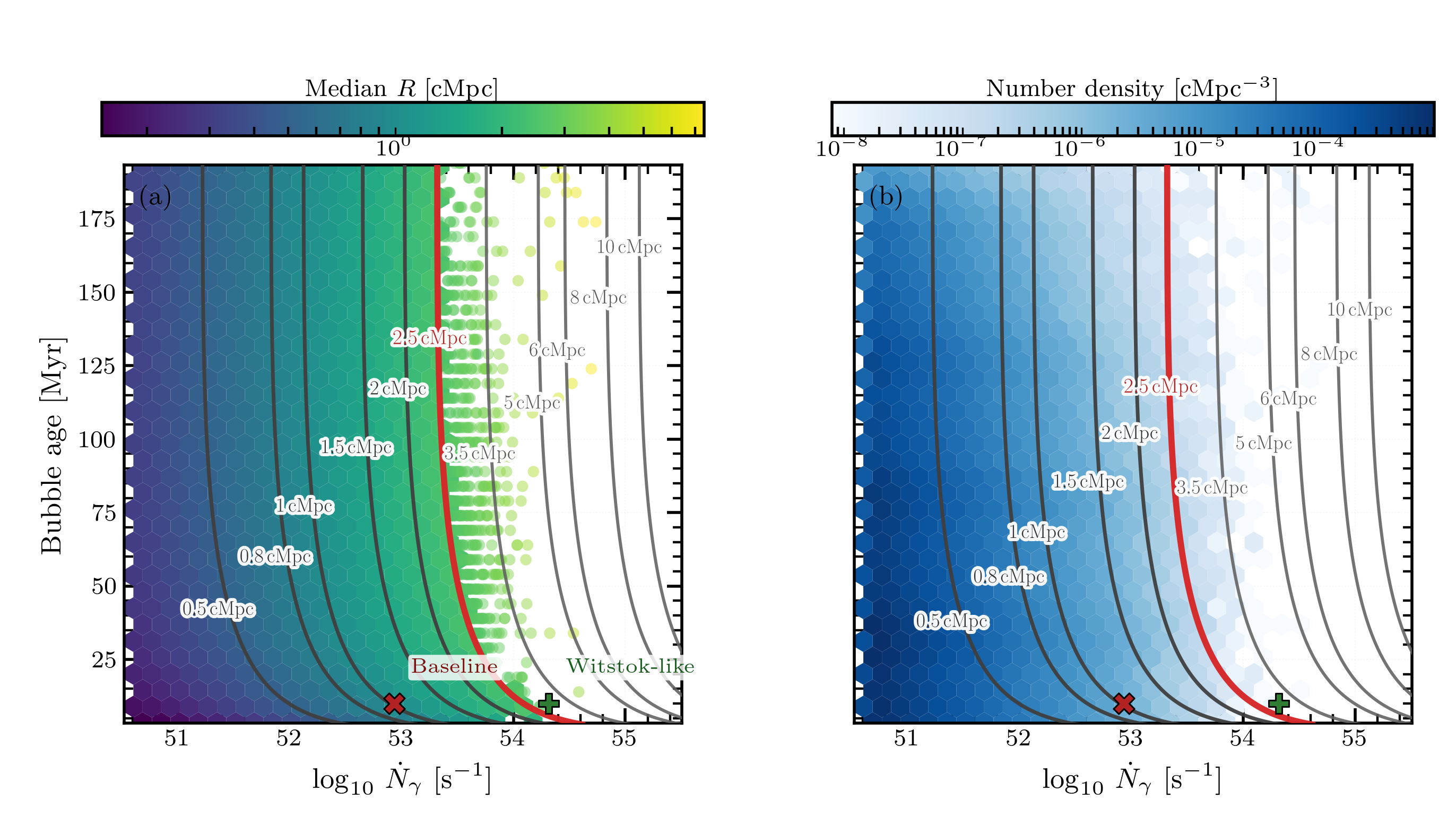}
    \caption{Non-merging source-group bubble age versus summed ionizing photon rate at $z=13$. The age is measured from the birth of the host bubble's first member. Panel~(a) shows the median radius in hexagonal bins for $R<2.5$~cMpc and individual bubbles for $R\ge2.5$~cMpc; panel~(b) shows the bubble-object number density in the same plane. Gray contours are recombination-aware constant-emissivity single-sphere references from Eq.~\eqref{eq:bubble_growth}, and the red contour marks $R=2.5$~cMpc. The red cross and green filled plus mark the baseline and adopted Witstok-like single-source reference parameters, respectively.}
    \label{fig:age_luminosity}
\end{figure*}

Figure~\ref{fig:age_luminosity} places this incidence in the $(\dot N_{\rm ion},\,\mathrm{age})$ plane of the same $z\approx 13$ bubble-object population.  Analytic contours in both panels (Appendix~\ref{app:analytic_bsd}) give constant-radius histories for one constant-emissivity sphere. Groups that acquire later members need not follow these reference curves. The $R\ge2.5$~cMpc tail is dominated by bright, old members: the brightest galaxy supplies a median $99.8\%$ of a group's integrated ionizing emission and is its oldest member in $99.3\%$ of cases. At $z=13$, evolving the same sources independently in the eight fiducial $L=250$~cMpc realizations gives 735 spheres above the threshold, compared with 743 source groups, a difference of $1.1\%$. The incidence therefore depends on the bright end of the UVLF and the duration of bright emission. This part of the population is especially sensitive to the assumption that member sources retain their assigned luminosities after birth. The activity experiments below interrupt emission to test its effect on the tail. Starting the evolution at $z=15.5$ lowers this surface density by $15.2\%$, illustrating its sensitivity to the extrapolated early source history. \S\ref{sec:discussion_witstok_object} illustrates the stellar and halo supply requirements for a bright source in this population.

\subsection{What controls the high-radius tail?}\label{sec:results_tail_controls}
Table~\ref{tab:sim_bubbles} shows how the $R\ge 2.5$~cMpc tail responds to variations in the UVLF and a small set of astrophysical parameters. Moderate parameter changes move the incidence by factors of a few, while larger shifts in the emissivity product can change it by orders of magnitude; the sparsest Bouwens/intermittent cases yield only upper limits. The comparison between the \citet{Donnan2024PRIMER} and \citet{Bouwens2021} determinations demonstrates strong UVLF-model dependence, but it does not isolate functional form alone because their fitted normalization, characteristic magnitude, slopes, and redshift evolution also differ.

\begin{deluxetable*}{lcccccccc}
\tabletypesize{\scriptsize}
\setlength{\tabcolsep}{4pt}
\tablecaption{Representative large-bubble statistics at \texorpdfstring{$z\!\approx\!13$}{z~13}\label{tab:sim_bubbles}}
\tablehead{
\colhead{Run} & \colhead{UVLF / mode} & \colhead{$M_{\rm lim}$} & \colhead{$f_{\mathrm{esc}}$} & \colhead{$\log \xi_{\mathrm{ion}}$} & \colhead{$f$} & \colhead{$C$} & \colhead{$n_{\ge2.5}^{\rm sim}$} & \colhead{$\Sigma_{\ge2.5}^{\rm sim}$}\\
\colhead{} & \colhead{} & \colhead{} & \colhead{} & \colhead{} & \colhead{} & \colhead{} & \colhead{[cMpc$^{-3}$]} & \colhead{[arcmin$^{-2}$ per $\Delta z=1$]} \\
}
\startdata
B21 comparison & B21 & $-13.0$ & 0.200 & 25.5 & 1.0 & 3 & $4.80^{+2.9}_{-1.9}\times10^{-8}$ & $6.83^{+1.7}_{-1.7}\times10^{-5}$ \\
B21 time duty & B21-T(10) & $-13.0$ & 0.200 & 25.5 & 0.2 & 3 & $8.00^{+18}_{-8}\times10^{-9}$ & $1.72^{+1.3}_{-1.3}\times10^{-5}$ \\
Low active fraction & D24 & $-13.0$ & 0.200 & 25.5 & 0.2 & 3 & $1.45^{+0.14}_{-0.13}\times10^{-6}$ & $1.97^{+0.17}_{-0.17}\times10^{-3}$ \\
D24 time duty & D24-T(10) & $-13.0$ & 0.200 & 25.5 & 0.2 & 3 & $2.08^{+0.12}_{-0.12}\times10^{-6}$ & $2.81^{+0.11}_{-0.11}\times10^{-3}$ \\
Intermediate active fraction & D24 & $-13.0$ & 0.200 & 25.5 & 0.5 & 3 & $2.92^{+0.18}_{-0.18}\times10^{-6}$ & $4.07^{+0.22}_{-0.22}\times10^{-3}$ \\
\textbf{Fiducial} & D24 & $\mathbf{-13.0}$ & \textbf{0.200} & \textbf{25.5} & \textbf{1.0} & \textbf{3} & $5.94^{+0.2}_{-0.2}\times10^{-6}$ & $8.10^{+0.25}_{-0.24}\times10^{-3}$ \\
Faint cutoff & D24 & $-12.0$ & 0.200 & 25.5 & 1.0 & 3 & $6.00^{+0.18}_{-0.18}\times10^{-6}$ & $8.20^{+0.25}_{-0.26}\times10^{-3}$ \\
Bright cutoff & D24 & $-14.0$ & 0.200 & 25.5 & 1.0 & 3 & $5.90^{+0.19}_{-0.18}\times10^{-6}$ & $8.11^{+0.25}_{-0.25}\times10^{-3}$ \\
Low clumping & D24 & $-13.0$ & 0.200 & 25.5 & 1.0 & 1 & $1.22^{+0.0096}_{-0.0096}\times10^{-5}$ & $1.66^{+0.011}_{-0.011}\times10^{-2}$ \\
High clumping & D24 & $-13.0$ & 0.200 & 25.5 & 1.0 & 20 & $2.96^{+0.4}_{-0.4}\times10^{-7}$ & $4.14^{+0.49}_{-0.49}\times10^{-4}$ \\
Low $f_{\rm esc}$ & D24 & $-13.0$ & 0.100 & 25.5 & 1.0 & 3 & $1.39^{+0.11}_{-0.11}\times10^{-6}$ & $1.90^{+0.13}_{-0.13}\times10^{-3}$ \\
High $f_{\rm esc}$ & D24 & $-13.0$ & 0.300 & 25.5 & 1.0 & 3 & $1.29^{+0.014}_{-0.014}\times10^{-5}$ & $1.76^{+0.013}_{-0.013}\times10^{-2}$ \\
Low $\xi_{\rm ion}$ & D24 & $-13.0$ & 0.200 & 25.0 & 1.0 & 3 & $4.64^{+0.4}_{-0.4}\times10^{-7}$ & $6.44^{+0.38}_{-0.38}\times10^{-4}$ \\
High $\xi_{\rm ion}$ & D24 & $-13.0$ & 0.200 & 26.0 & 1.0 & 3 & $4.69^{+0.021}_{-0.021}\times10^{-5}$ & $6.41^{+0.019}_{-0.019}\times10^{-2}$ \\
\enddata
\tablecomments{Compact subset of the parameter study. Unsuffixed UVLF labels use the active-fraction model and $f=f_{\rm active}$; T($t_{\rm on}$) marks the time-duty model and $f=f_{\rm time}$, with $t_{\rm on}$ in Myr. Each row uses the eight large-volume runs described in \S\ref{sec:model}; the $C=20$ rare-tail run uses refined growth intervals (Table~\ref{tab:sim_bubbles_full}). Simulation number densities use the exact $z=13.0$ output; simulation surface densities use the five-slice $\Delta z=1$ estimator. Error bars give $1\sigma$ box-resampling intervals, supplemented by stacked-count Garwood bounds for sparse cases. Error bars do not include large-scale-structure cosmic variance, UVLF uncertainty, or model uncertainty. Labels are defined in Table~\ref{tab:params}; the full parameter study is in Appendix~\ref{app:full_parameter_suite}.}
\end{deluxetable*}

Table~\ref{tab:sim_bubbles} compresses the parameter variations to two incidence statistics (number density and sky surface density), while \S\ref{sec:discussion_sensitivity} examines the same trends in the full BSDs. The main trend is that parameters which change the emissivity of already-bright sources move the high-radius tail most efficiently.

\subsection{Global ionization history and morphology}\label{sec:results_global_morphology}

\begin{figure*}[!t]
    \includegraphics[width=\textwidth]{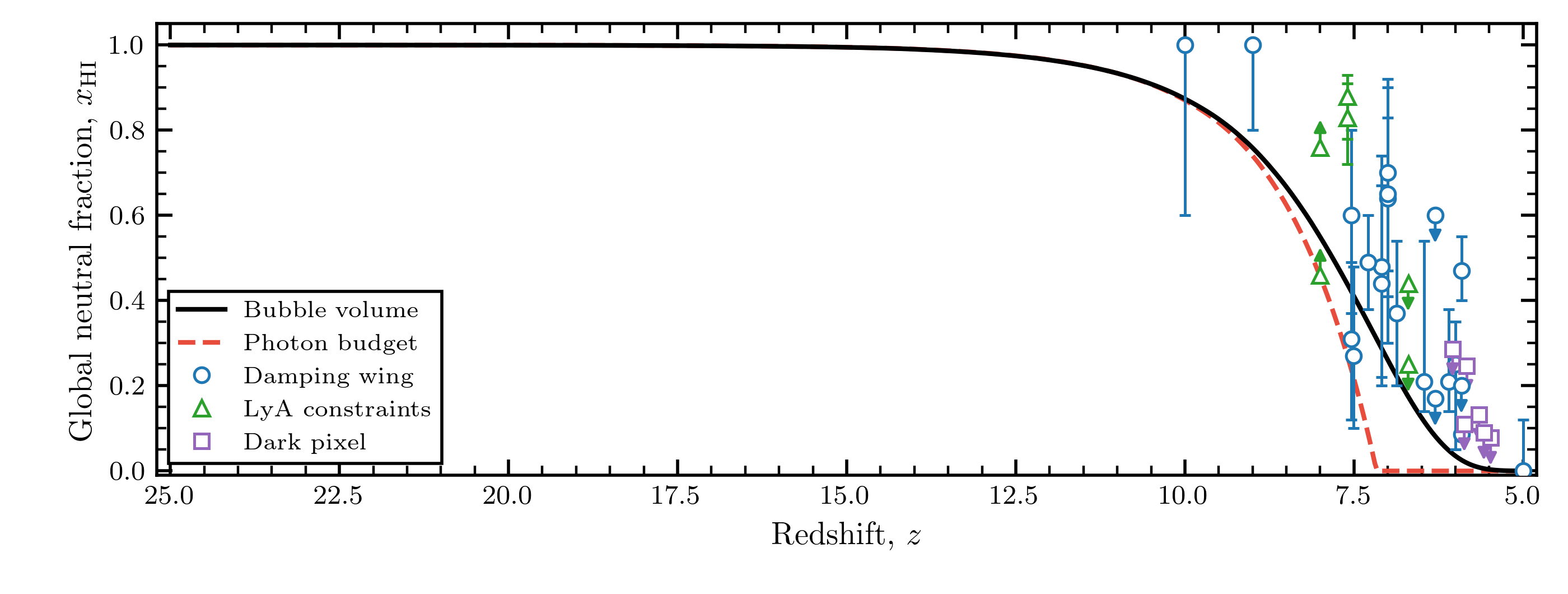}
    \caption{Global neutral hydrogen fraction as a function of redshift. The solid black curve shows the painted-filling estimate and the dashed red curve the photon-budget estimate. Observational constraints are grouped by probe: damping wings in GRBs, individual and stacked quasars, and galaxies \citep{Totani2006,Totani2014,Fausey2025,Davies2018,Wang2020,Greig2022,Durovcikova2024,Umeda2025}; Ly$\alpha$ emitter fractions and equivalent widths \citep{Hoag2019,Mason2019,Bolan2022}; and dark pixels in the Ly$\alpha$, Ly$\beta$, and Ly$\gamma$ forests \citep{McGreer2015,Davies2026}. Arrows point from the reported bounds toward allowed values. Bounds from \citet{Totani2006,Mason2019,Bolan2022} show 68\% and 95\% levels; those from \citet{McGreer2015,Davies2026} are $1\sigma$. The \citet{Fausey2025} bound is $3\sigma$ for the McQuinn model with a Gaussian spectral-index prior. The \citet{Davies2026} points use negative-pixel Ly$\beta$+Ly$\gamma$ estimates; stacked-quasar points use ATON results. Some samples overlap, and the constraints are shown for comparison only.}
    \label{fig:ion_prog}
\end{figure*}

Figure~\ref{fig:ion_prog} shows the neutral fraction, $x_{\mathrm{HI}}$, from two estimates: the occupied fraction of the sphere-painted grid and the cumulative photon budget distributed homogeneously. Continuing the Donnan DPL formula to low redshift gives a predominantly neutral universe at early times, a rapid transition over $6\lesssim z\lesssim8$, and a highly ionized state by $z\simeq6$.

The estimates agree at high redshift but separate during the transition, with a maximum difference of approximately 0.31 near $z=7.15$. The photon-budget estimate distributes ionization homogeneously; the painted estimate caps overlapping occupancies at unity and leaves cells outside the spheres neutral.

The painted and photon-budget histories reach $x_{\rm HI}=0.5$ at $z\simeq7.82$ and 8.11, respectively, within the broad observational range \citep{Greig2022,Durovcikova2024,Mason2019,Bolan2022}. The painted history reaches $x_{\rm HI}<0.01$ by $z\simeq5.67$ \citep[cf.][]{Bosman2022}. These comparisons depend on the low-redshift UVLF continuation and are not used to calibrate the $z\approx13$ incidence. A CMB optical-depth comparison would additionally require helium ionization and a low-redshift electron history \citep{Planck2018,Pagano2020}. At $z\simeq13$, $x_{\rm HI}\simeq0.981$: large spheres are present while most of the volume remains neutral.

\begin{figure*}[!t]
    \centering
    \includegraphics[width=0.96\textwidth]{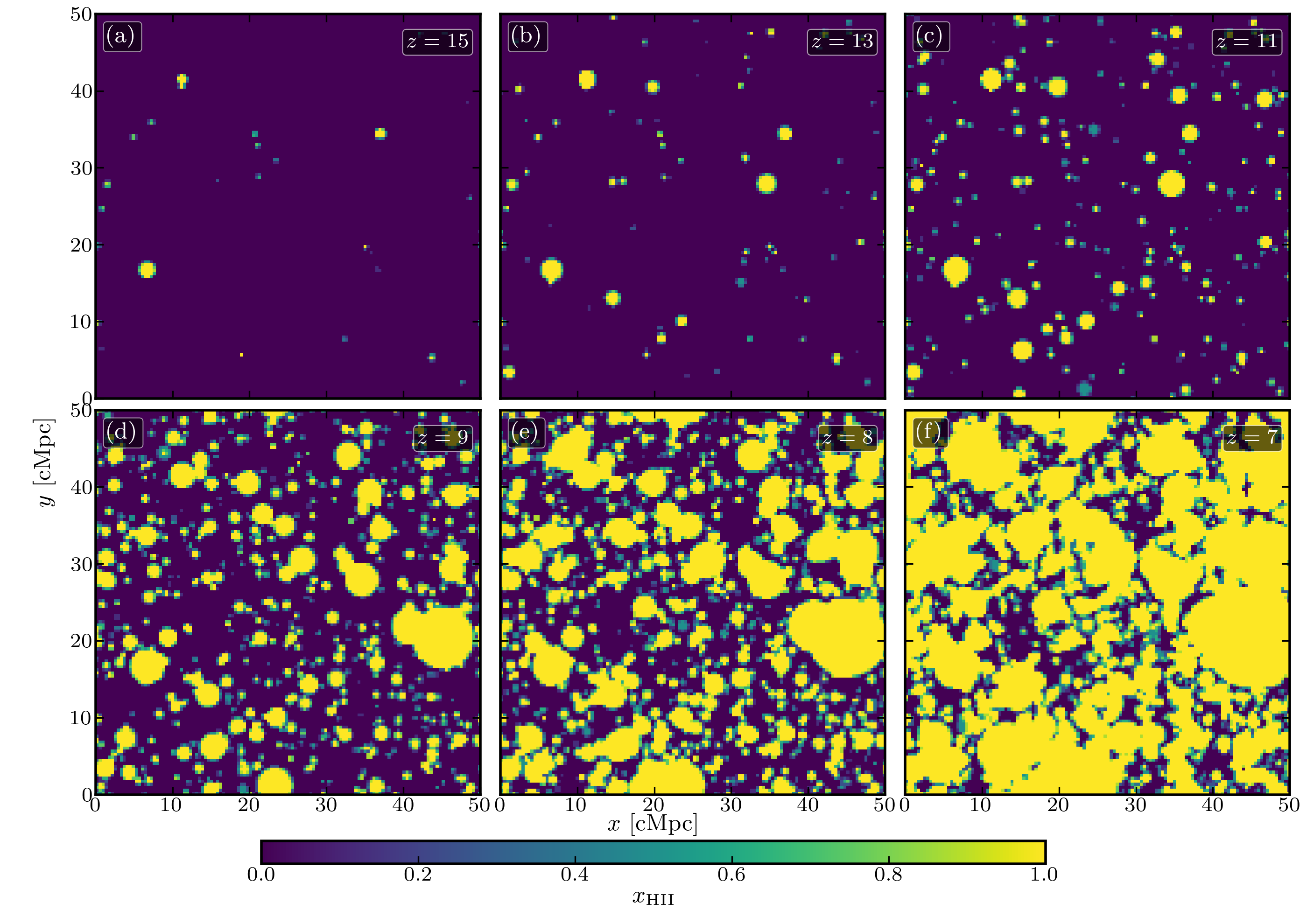}
    \caption{One-cell-thick central slices through the painted-sphere filling field of the fiducial $L=50$~cMpc, $128^3$ run (cell width 0.39~cMpc). From top-left to bottom-right, redshifts decrease as \texorpdfstring{$z=15,13,11,9,8,7$}{z=15,13,11,9,8,7}. Color gives the subvoxel ionized occupancy $x_{\rm HII}$ after overlapping sphere contributions are capped at unity.}
    \label{fig:slice}
\end{figure*}

The painted filling field evolves substantially throughout reionization. At early times, neutral cells are punctuated by isolated, approximately spherical ionized regions. With decreasing redshift, the spheres expand and overlap, leaving neutral gaps and interstices between connected painted regions. Figure~\ref{fig:slice} visualizes this geometric occupancy only; it does not contain the density fluctuations or anisotropic transfer needed to identify physical neutral filaments.

\subsection{BSD evolution and analytic comparison}\label{sec:results_bsd_evolution}

\begin{figure*}[!ht]
    \centering
    \includegraphics[width=0.96\textwidth]{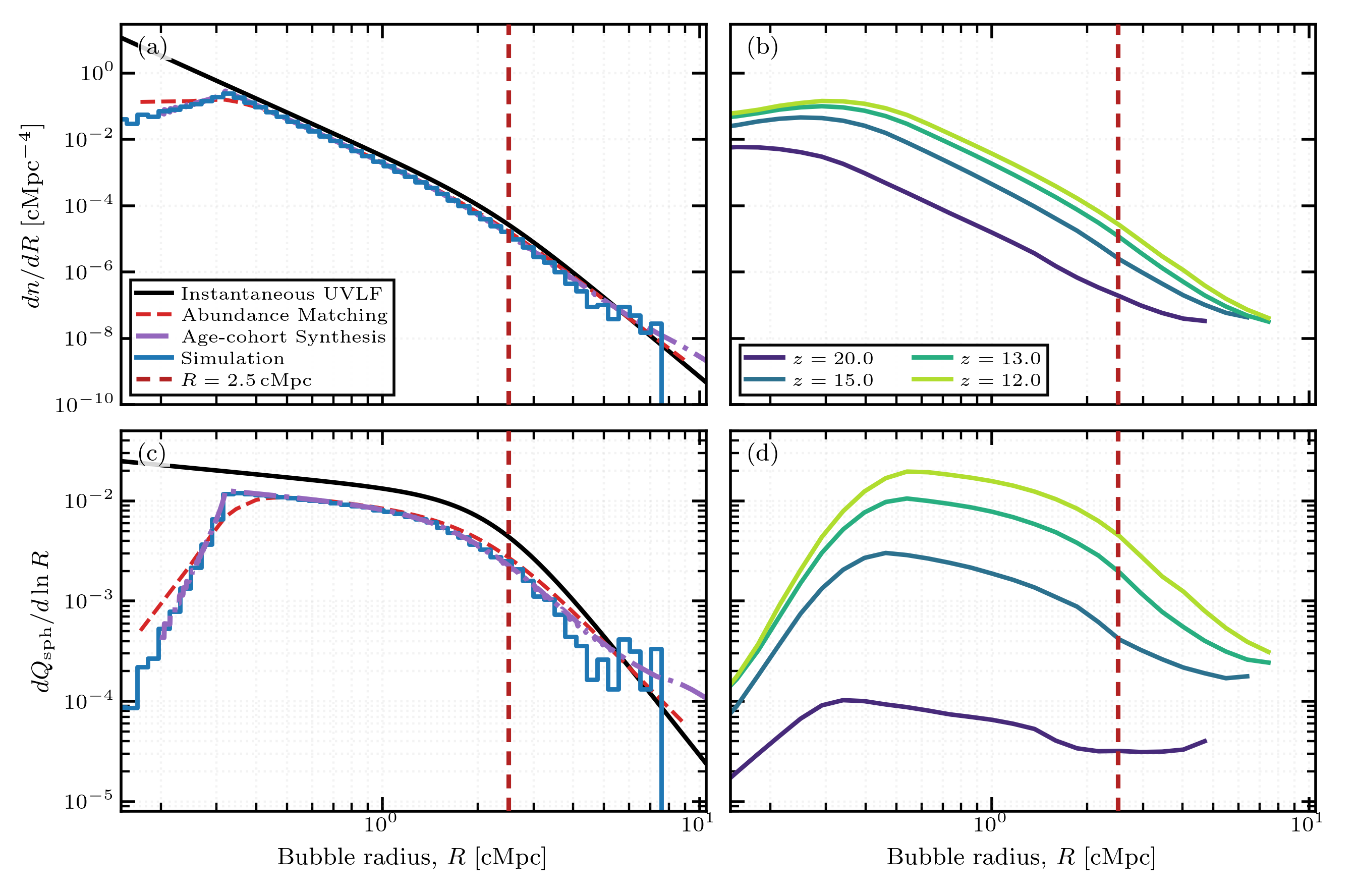}
    \caption{BSD and sphere-volume contributions from the eight $L=250$~cMpc realizations. Panel~(a) compares the $z=13$ simulation BSD with the UVLF-based approximate analytic mappings in Appendix~\ref{app:analytic_bsd}; panel~(c) weights each non-merging source-group sphere by its volume. Panels~(b) and~(d) show the redshift evolution of these quantities. $Q_{\rm sph}$ sums sphere volumes, including geometric overlaps. The red line marks $R=2.5$~cMpc.}
    \label{fig:hist_z13}
\end{figure*}

Figure~\ref{fig:hist_z13} collects the BSD and its volume contribution. Panel~(a) compares the $z\approx 13$ bubble size distribution with the three analytic prescriptions derived in Appendix~\ref{app:analytic_bsd}: Method~1 maps the instantaneous UVLF onto the BSD through an effective photon-weighted timescale; Method~2 synthesizes the BSD from age-cohort birth-rate integrals with a recombination-weighted lifetime and then applies a Poisson approximation to the baseline source-attachment multiplicity; Method~3 evolves rank-ordered sources through the ionizing-photon budget directly. The plotted Method~1 curve adopts an illustrative 50~Myr effective duration; changing this duration shifts its radius scale, so it is neither a steady-state solution nor an absolute bound on the tail. Methods~2 and~3 both track the high-$R$ tail more accurately because they integrate the recombination-limited emissivity history. A multiplicity-resolved decomposition appears in Appendix~\ref{app:analytic_bsd}.

Panel~(c) shows the corresponding sphere-object volume contribution, $\mathrm{d}Q_{\rm sph}/\mathrm{d}\ln R$. Large bubbles are rare by number, but their volumes scale as $R^3$, so they contribute disproportionately to this sum. Overlapping source groups contribute separately to this sum. 

Panels~(b) and~(d) extend the same two quantities across redshift. At the earliest displayed epochs the distribution is dominated by small bubbles around recently formed sources. As cosmic time advances, the distribution broadens to larger radii. The $z\approx 13$ result measures the high-radius tail while the typical number-weighted bubble remains small.

\section{Discussion}\label{sec:discussion}

\begin{figure*}[!t]
    \centering
    \includegraphics[width=0.96\textwidth]{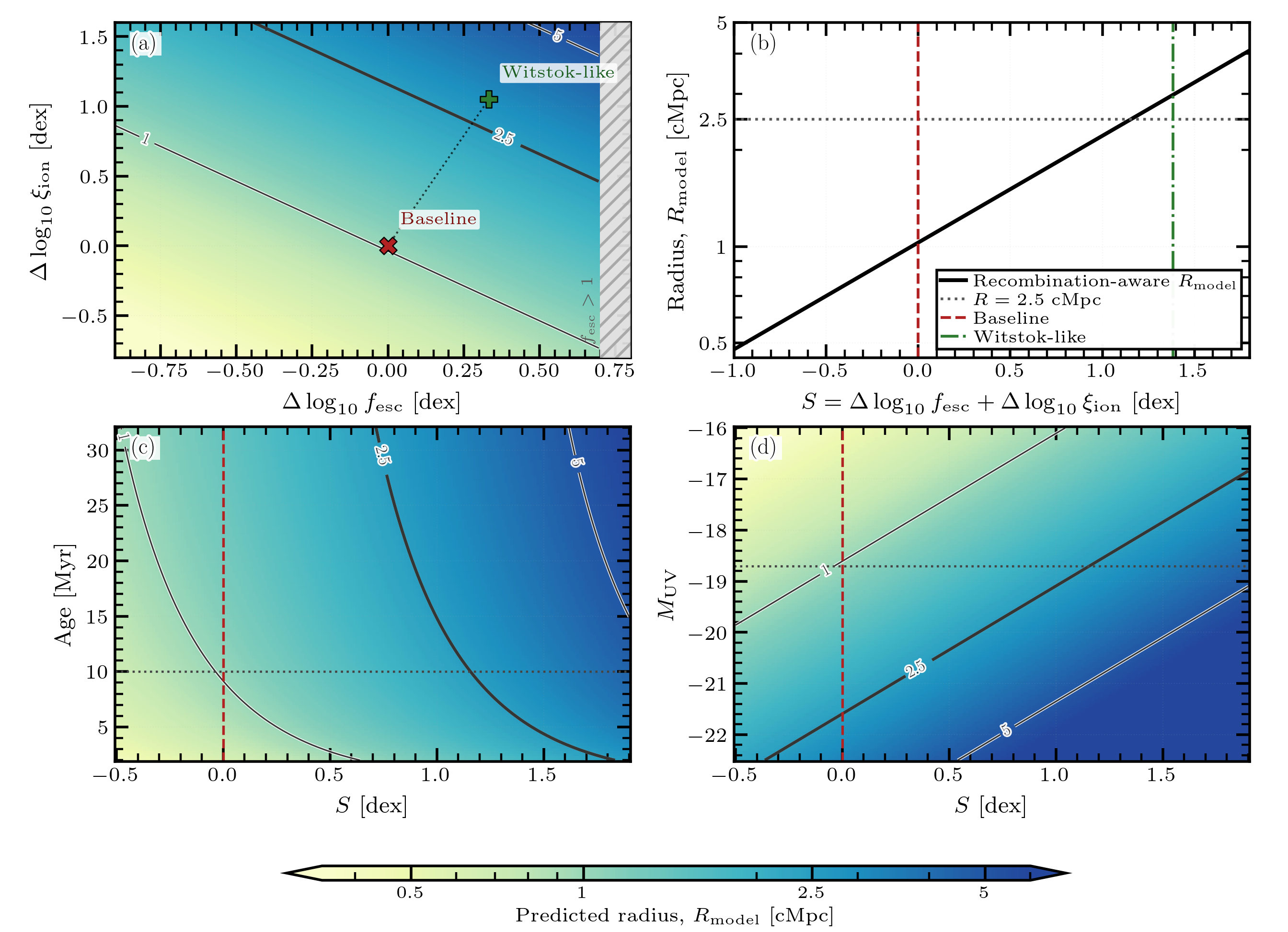}
    \caption{Deterministic source-level radius from the recombination-aware growth law at $z=13$ and $C=3$. Panel~(a) varies $f_{\rm esc}$ and $\xi_{\rm ion}$ at $M_{\rm UV}=-18.7$ and age 10~Myr; the hatched region has $f_{\rm esc}>1$. Panel~(b) shows the same radius against $S=\Delta\log_{10}(f_{\rm esc}\xi_{\rm ion})$, while panels~(c) and~(d) additionally vary age and $M_{\rm UV}$.}
    \label{fig:source_level_degeneracy}
\end{figure*}

\subsection{From population incidence to the Witstok object}\label{sec:discussion_witstok_object}

The inferred ionized region around JADES-GS-z13-1-LA is large for its observed $M_{\rm UV}\simeq-18.7$ \citep{Witstok2025Nature}. Although our fiducial population produces regions of this size, their sources can be brighter or older. We compare the observed source with the growth law using the emissivity shift $S=\Delta\log_{10}(f_{\rm esc}\xi_{\rm ion})$ at fixed UV luminosity and age.

Figure~\ref{fig:source_level_degeneracy} summarizes the source-level degeneracy using the recombination-aware radius from Eq.~\eqref{eq:bubble_growth}. This is a fixed-parameter illustration. We fix $z=13$, $C=3$, and, except where varied, $M_{\rm UV}=-18.7$ and a 10~Myr source age. Panels~(a) and~(b) map the radius in the $(\Delta\log f_{\rm esc},\Delta\log\xi_{\rm ion})$ plane and against the combined emissivity shift $S$. The diagonal structure in panel~(a) reflects the dependence on the product $f_{\rm esc}\xi_{\rm ion}$. The region with $f_{\rm esc}>1$ is masked.

The Witstok reference adopts $f_{\rm esc}=0.43$ and $\log_{10}\xi_{\rm ion}=26.55$ from \citet{Witstok2025Nature}. At fixed UV luminosity and age, these values give an emissivity shift $S=1.38$~dex relative to our fiducial parameters. 

\begin{figure*}[!t]
    \setlength{\abovecaptionskip}{1pt}\setlength{\belowcaptionskip}{0pt}
    \centering
    \includegraphics[width=0.96\textwidth]{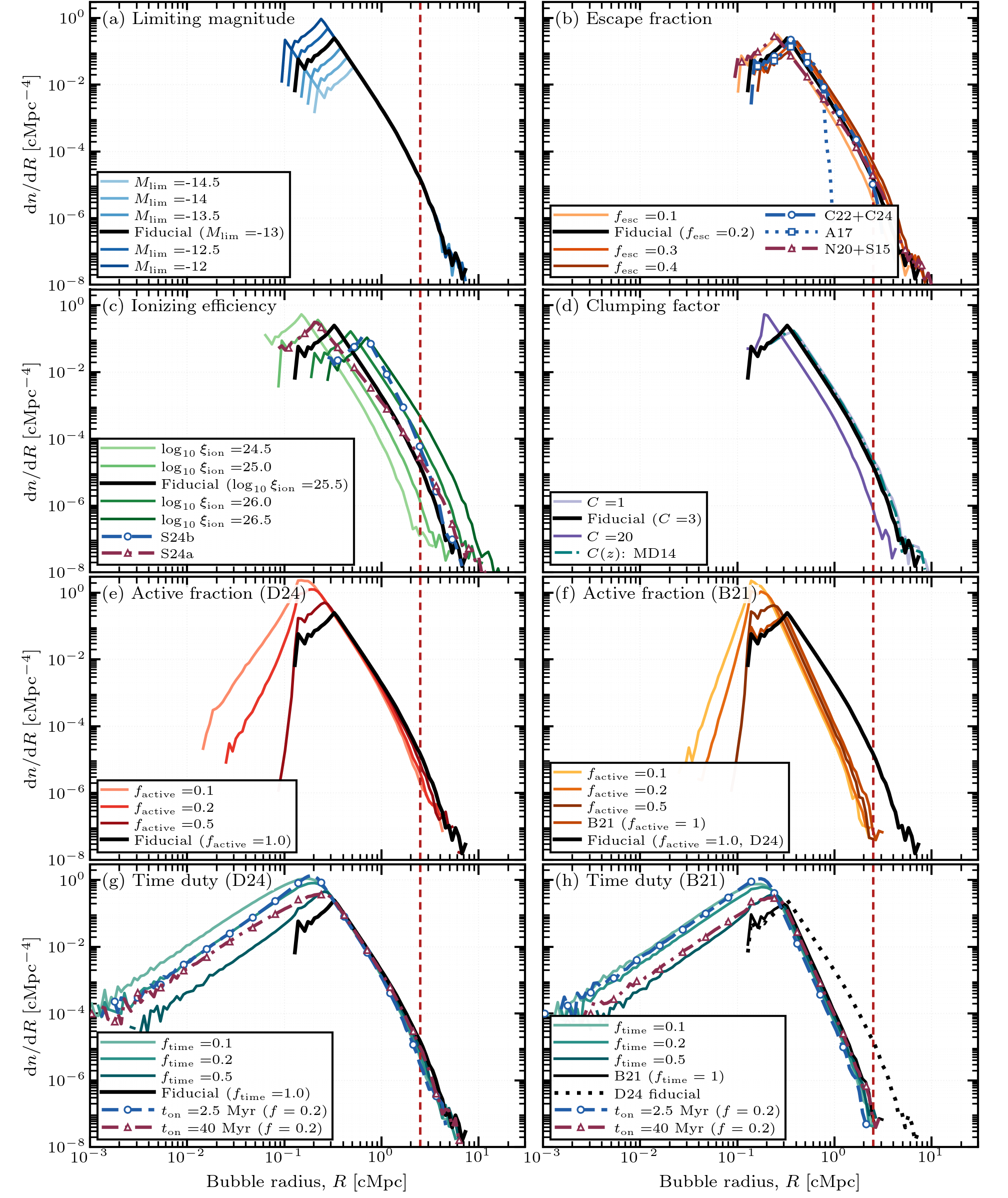}
    \caption{BSDs at $z=13$ from eight $L=250$~cMpc boxes per case. Panels (a--d) vary $M_{\rm lim}$, $f_{\rm esc}$, $\xi_{\rm ion}$, and $C$; (e,f) vary the active fraction. In (g,h), teal shading varies $f_{\rm time}$ at $\langle t_{\rm on}\rangle=10$~Myr; colored broken lines vary the mean on-time at $f_{\rm time}=0.2$. All panels use identical bins and normalization. Labels are defined in Table~\ref{tab:params}; symbols identify curves. The red line marks $R=2.5$~cMpc.}
    \label{fig:bsd_group_comparison}
\end{figure*}

Panels~(c) and~(d) show how the deterministic radius responds to source age and observed UV magnitude at fixed values of the other parameters. Figure~\ref{fig:bsd_group_comparison} separately shows population-level variations in the UVLF cutoff and clumping factor.

\begin{figure*}[!t]
    \centering
    \includegraphics[width=0.96\textwidth]{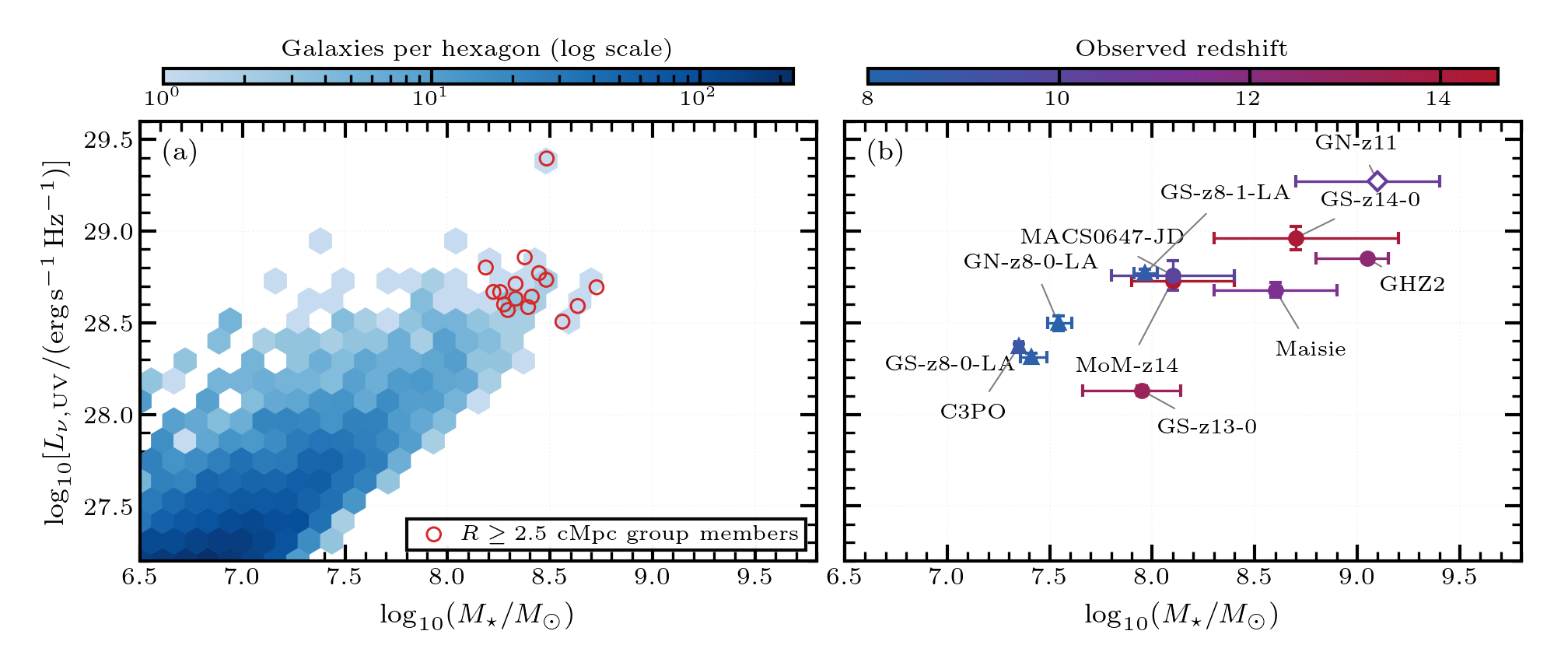}
    \caption{UV luminosity and stellar mass. (a) Reconstructed surviving Chabrier masses at $z=13$ in four $L=100$~cMpc fiducial boxes. Hexagons show galaxy counts; red circles mark individual members of $R\ge2.5$~cMpc groups. Of 184,967 galaxies, 4,071 lie within the displayed range, including 16 large-group members. These are a separate mass-history sample from the eight $L=250$~cMpc incidence boxes. (b) Eleven selected $8<z<14.5$ galaxies, colored by observed redshift. Circles show MoM-z14 \citep{Naidu2026MoM}, GS-z14-0 \citep{Schouws2025GSz14}, GS-z13-0 \citep{CurtisLake2023}, GHZ2 \citep{Castellano2024GHZ2}, Maisie's Galaxy \citep{ArrabalHaro2023}, and MACS0647-JD \citep{Hsiao2024NIRSpec,Hsiao2024MIRI}. Triangles show JADES-GN-z8-0-LA, JADES-GS-z8-0-LA, and JADES-GS-z8-1-LA \citep[][Table~2, NIRCam UV magnitudes]{Witstok2025JADESLAEs}, and C3PO/CEERS-24 \citep[][Table~2, adopted constant-SFH fit]{Hu2026C3PO}. JADES names are abbreviated in the plot. The open diamond marks GN-z11's total-galaxy fit assuming no AGN light \citep{Tacchella2023GNz11}. GHZ2 uses the published BAGPIPES fit; GHZ2 and MACS0647-JD include the cited lensing corrections. Luminosities are not uniformly corrected for dust. Error bars retain published marginal uncertainties and exclude covariance and additional IMF, star-formation-history, and lensing systematics.}
    \label{fig:highz_mass_uv}
\end{figure*}

Possible source-level explanations include the following.

A recent burst can generate an ionized region that persists as the UV continuum fades. Bursty histories are supported by \textit{JWST} observations and modeling \citep{Endsley2023JADES,Endsley2024Burstiness,Stiavelli2026,Munoz2026Burstiness}; the mini-quenched galaxy at $z\approx7.3$ provides an example of rapid fading \citep{Looser2024}. The bubble then reflects the brighter past, reducing the instantaneous $f_{\rm esc}\xi_{\rm ion}$ needed to explain its size.

The system may have a high escape fraction or unusually efficient stellar ionizing production. Figure~\ref{fig:source_level_degeneracy} shows the required increase in their product for the adopted age and luminosity. Distinguishing this possibility from an earlier burst or AGN contribution requires additional constraints on the source spectrum and history.

An accreting black hole can harden the spectrum and increase the ionizing output per unit observed UV continuum. Broad-line AGN among compact red sources at high redshift provide evidence for this channel \citep{Greene2024}. An AGN contribution could raise the effective $\xi_{\rm ion}$ of the Witstok source beyond the stellar value assumed in our fiducial calculation \citep{Cohon2026LAE}.

Figure~\ref{fig:highz_mass_uv}(b) collects stellar-mass and UV measurements for eleven selected $8<z<14.5$ galaxies, including the other Ly$\alpha$ emitters discussed here. GN-z11 shows evidence for an accreting black hole \citep{Bunker2024GNz11,Maiolino2024GNz11}, while the three JADES $z\approx8$ emitters and C3PO/CEERS-24 at $z=8.998$ are dominated by recent star formation \citep{Witstok2025JADESLAEs,Hu2026C3PO}. The latter has an inferred stellar age of $1.6\pm0.2$~Myr. These source-history constraints and the mass--UV measurements provide complementary context, but neither directly measures the duration of ionizing emission that built a surrounding bubble. Earlier episodes and neighboring sources can contribute, and predicting transmission for the Witstok benchmark requires radiative transfer along the observed sight line \citep{Qin2025Lya}.

Figure~\ref{fig:highz_mass_uv}(a) shows surviving stellar masses reconstructed at $z=13$ from four $L=100$~cMpc fiducial realizations. We integrate UV-inferred star formation rates from source birth with zero initial stellar mass, adopting a Chabrier IMF \citep{MadauDickinson2014} and age-dependent mass return \citep{Shuntov2025}. These masses do not feed back on the bubble calculation. At fixed luminosity, greater mass reflects a longer emitting history. The mixed-redshift catalog in panel~(b) provides context but cannot validate the $z=13$ stellar mass function because its selection and stellar-population assumptions differ.

\paragraph{Stellar mass and host supply.}\label{sec:host_budget}
At the Witstok luminosity, $M_{\rm UV}=-18.7$, a single source with fiducial efficiencies reaches only $R\simeq1.84$~cMpc by $z=13$ even if continuously active from $z=25$. We therefore illustrate the supply requirement for a brighter member of the simulated population: an $M_{\rm UV}=-20$ source reaches $R=2.5$~cMpc after approximately 65~Myr. The Salpeter UV--SFR conversion of \citet{MadauDickinson2014} gives ${\rm SFR}\simeq5~M_\odot\,{\rm yr}^{-1}$ and $3.25\times10^8~M_\odot$ of formed stars.

Abundance matching the adopted $M_{\rm UV}\le-20$ density, $8.0\times10^{-6}$~cMpc$^{-3}$, to the \citet{ShethTormen1999} halo mass function gives $M_h\simeq2.2\times10^{10}~M_\odot$, assuming one galaxy per halo and no luminosity scatter. The \citet{Correa2015AccretionII} mean accretion rate implies gross baryon inflow of approximately $72~M_\odot\,{\rm yr}^{-1}$ at $z=13$. This exceeds the required SFR, but extrapolation and uncertain gas delivery prevent it from establishing sustained emission for 65~Myr.

\begin{figure*}[!t]
    \centering
    \includegraphics[width=0.96\textwidth]{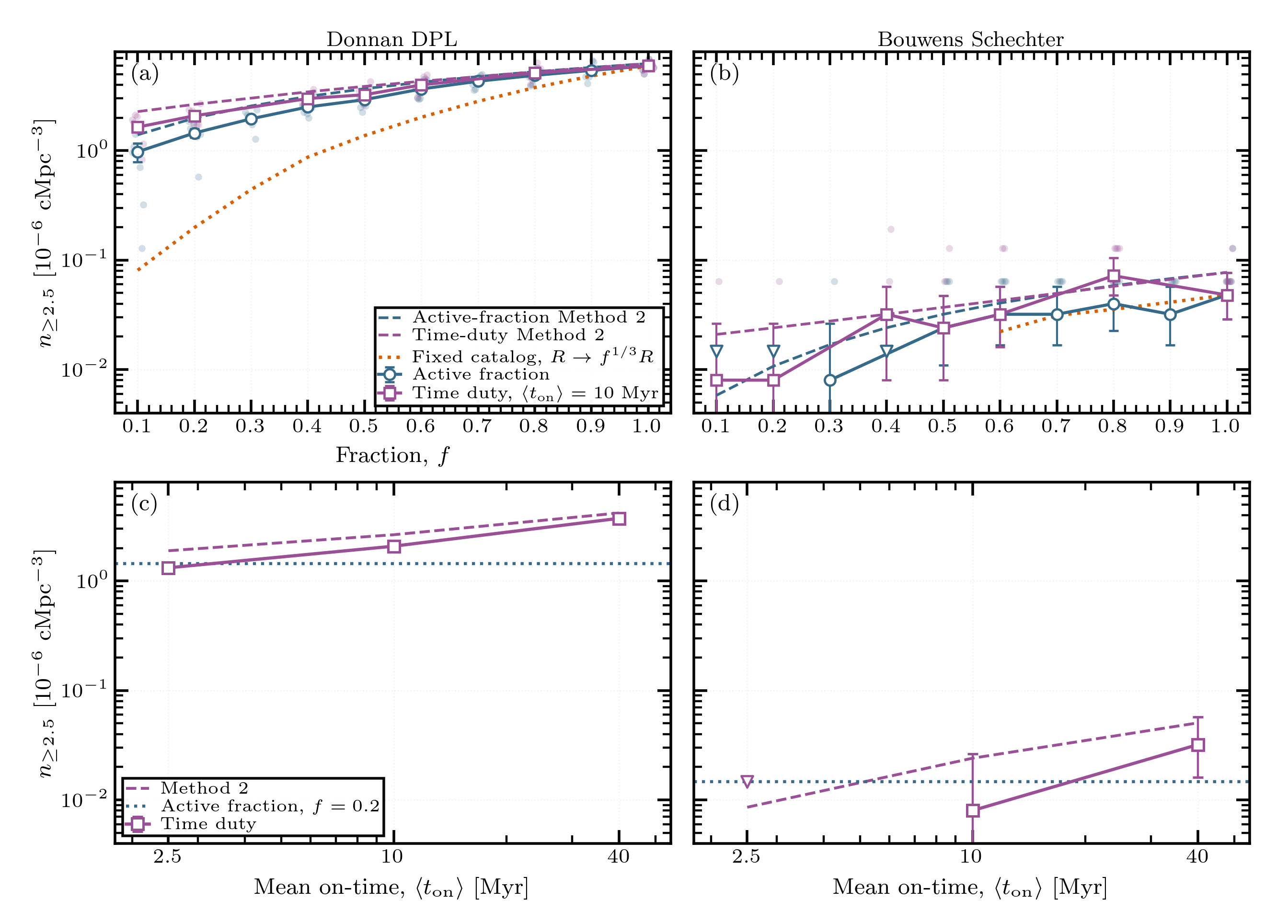}
    \caption{UVLF-calibrated source-activity effects at $z=13$. Panels~(a) and~(b) compare active-fraction and time-duty incidence for the D24 and B21 UVLFs with approximate Method-2 comparisons and the fixed-catalog cube-root reference. Panels~(c) and~(d) vary mean on-time at $f_{\rm time}=0.2$. Small points show the eight $L=250$~cMpc realization values.}
    \label{fig:active_fraction_incidence}
\end{figure*}

\subsection{Physical scalings behind the parameter trends}\label{sec:discussion_sensitivity}
We interpret the incidence trends in Section~\ref{sec:results_tail_controls} in terms of the adopted UVLF, ionizing efficiencies, and source-activity histories, and examine their effects on the full BSD.

The Donnan double-power-law UVLF has a more populated bright end than the Bouwens Schechter model, favoring large bubbles; the comparison also changes the fitted parameters and redshift evolution (\S\ref{sec:results_tail_controls}). Changing the faint-end cutoff has a weaker effect on the tail [Fig.~\ref{fig:bsd_group_comparison}(a)].

At fixed source history, the growth law gives $R\propto f_{\rm esc}^{1/3}$ (\S\ref{sec:param_degeneracies}). Raising $f_{\rm esc}$ from 0.2 to 0.3 therefore increases the radius by a factor of 1.14, shifting the BSD horizontally in Fig.~\ref{fig:bsd_group_comparison}(b). The time to reach a fixed radius instead scales as $f_{\rm esc}^{-1}$ in the photon-counting limit; recombinations can prevent the target from being reached.

We test luminosity-dependent escape directly in Fig.~\ref{fig:bsd_group_comparison}(b), applying each prescription to individual simulated galaxies while retaining the fiducial source population and histories. Table~\ref{tab:params} identifies the prescriptions. The escape prescriptions of \citet{Chisholm2022} and \citet{Naidu2020}, combined with the UV-slope and size relations of \citet{Cullen2024} and \citet{Shibuya2015}, respectively, give $n_{\ge2.5}=2.01\times10^{-6}$ and $9.50\times10^{-6}$~cMpc$^{-3}$, respectively, compared with $5.94\times10^{-6}$~cMpc$^{-3}$ for constant $f_{\rm esc}=0.2$. The \citet{Anderson2017} prescription yields no threshold objects in the simulated volume, corresponding to a one-sided 95\% Poisson upper limit of $2.40\times10^{-8}$~cMpc$^{-3}$. Thus the high-radius tail can be suppressed or enhanced by the assumed luminosity dependence. These comparisons change both the total photon supply and its distribution among galaxies; they do not isolate either contribution.

Figure~\ref{fig:bsd_group_comparison}(c) similarly tests two $\xi_{\rm ion}(M_{\rm UV})$ prescriptions at fixed $f_{\rm esc}=0.2$: the emission-line-selected fit of \citet{Simmonds2024} and the mass-complete sample's star-forming relation of \citet{Simmonds2024MassComplete}. They give $n_{\ge2.5}=1.85\times10^{-5}$ and $1.48\times10^{-5}$~cMpc$^{-3}$, respectively. All five tests extrapolate lower-redshift relations without intrinsic scatter; they illustrate sensitivity rather than establish calibrated efficiencies at $z=13$. Table~\ref{tab:sim_bubbles_full} includes their incidence statistics.

Larger effective clumping factors $C$ suppress growth at late times by shortening $t_{\rm rec}$ as defined in Eq.~\eqref{eq:t_rec} and reducing $R_{\rm S}$. This preferentially reduces the abundance of intermediate- and large-$R$ bubbles and narrows the distribution, as shown in Fig.~\ref{fig:bsd_group_comparison}(d).

The activity prescriptions in \S\ref{sec:model} separate population occupancy, $f_{\rm active}$, from the stationary fraction and episode duration of a correlated source history. They test the effect of intermittent emission at fixed on-state luminosity, motivated by evidence for bursty histories in early galaxies \citep{Witstok2025JADESLAEs,Endsley2023JADES,Endsley2024Burstiness,Stiavelli2026,Munoz2026Burstiness}. Appendix~\ref{app:method2} describes the approximate analytic treatment.

The clustering analysis of \citet{Dalmasso2024Clustering} infers a halo-occupation duty factor $\epsilon_{\rm DC}=0.075^{+0.540}_{-0.067}$ for its highest-redshift bin, whose mean redshift is $z=10.6$. This quantity comes from the relative abundances of UV-selected galaxies and halos and does not measure individual light curves. Of our two phenomenological controls, $f_{\rm active}$ is the closer comparison, although the quantities are not identical. The nearest directly simulated model, $f_{\rm active}=0.1$, gives $n_{\ge2.5}=9.76\times10^{-7}$~cMpc$^{-3}$ and $\Sigma_{\ge2.5}=1.36\times10^{-3}$~arcmin$^{-2}$, approximately one-sixth of the continuously active fiducial incidence. At the central value $f_{\rm active}=0.075$, for which we did not run a separate simulation, the Method-2 approximation predicts $n_{\ge2.5}=1.25\times10^{-6}$~cMpc$^{-3}$ and $\Sigma_{\ge2.5}=1.70\times10^{-3}$~arcmin$^{-2}$ per $\Delta z=1$. Method 2 predicts higher low-activity incidences than the simulations (Table~\ref{tab:sim_bubbles_full}).

Figure~\ref{fig:active_fraction_incidence} separates population occupancy from temporal coherence in the high-radius tail. Panels~(a) and~(b) show strong suppression toward lower activity fractions in the Donnan model; the much sparser Bouwens counts constrain this trend less precisely. The fixed-catalog cube-root mapping gives a stronger suppression because it reduces the integrated output of a fixed source catalog; the simulated models recalibrate the emitting population to the UVLF. The approximate Method-2 curves provide scaling comparisons, subject to the activity-kernel and companion-volume approximations described in Appendix~\ref{app:analytic_bsd}. At $f=0.2$ in the Donnan model, $\Sigma_{\ge2.5}$ decreases from $8.10\times10^{-3}$ to $1.97\times10^{-3}$~arcmin$^{-2}$ for instantaneous occupancy and to $2.81\times10^{-3}$~arcmin$^{-2}$ for 10~Myr time-duty episodes. The much lower Bouwens incidence persists under both prescriptions, separating UVLF-model dependence from source temporal behavior.

Panels~(c) and~(d) hold the stationary time-duty fraction fixed at $f_{\rm time}=0.2$ and vary $\langle t_{\rm on}\rangle$. Longer correlated episodes broaden the distribution of cumulative source activity and strengthen the rare large-$R$ tail without changing the mean emissivity budget; shorter episodes approach more rapid temporal averaging. Across the Donnan on-time tests, the escaped-photon budgets summed over the eight realizations through $z=13$ agree within $0.2\%$, supporting a temporal-history origin for the much larger changes in threshold counts. For the Donnan UVLF, $n_{\ge2.5}$ rises from $1.32\times10^{-6}$ to $3.74\times10^{-6}$~cMpc$^{-3}$ as $\langle t_{\rm on}\rangle$ increases from 2.5 to 40~Myr; the Bouwens comparison remains sparse enough that zero-count cases are reported as upper limits.

\begin{figure*}[!t]
    \centering
    \includegraphics[width=0.96\textwidth]{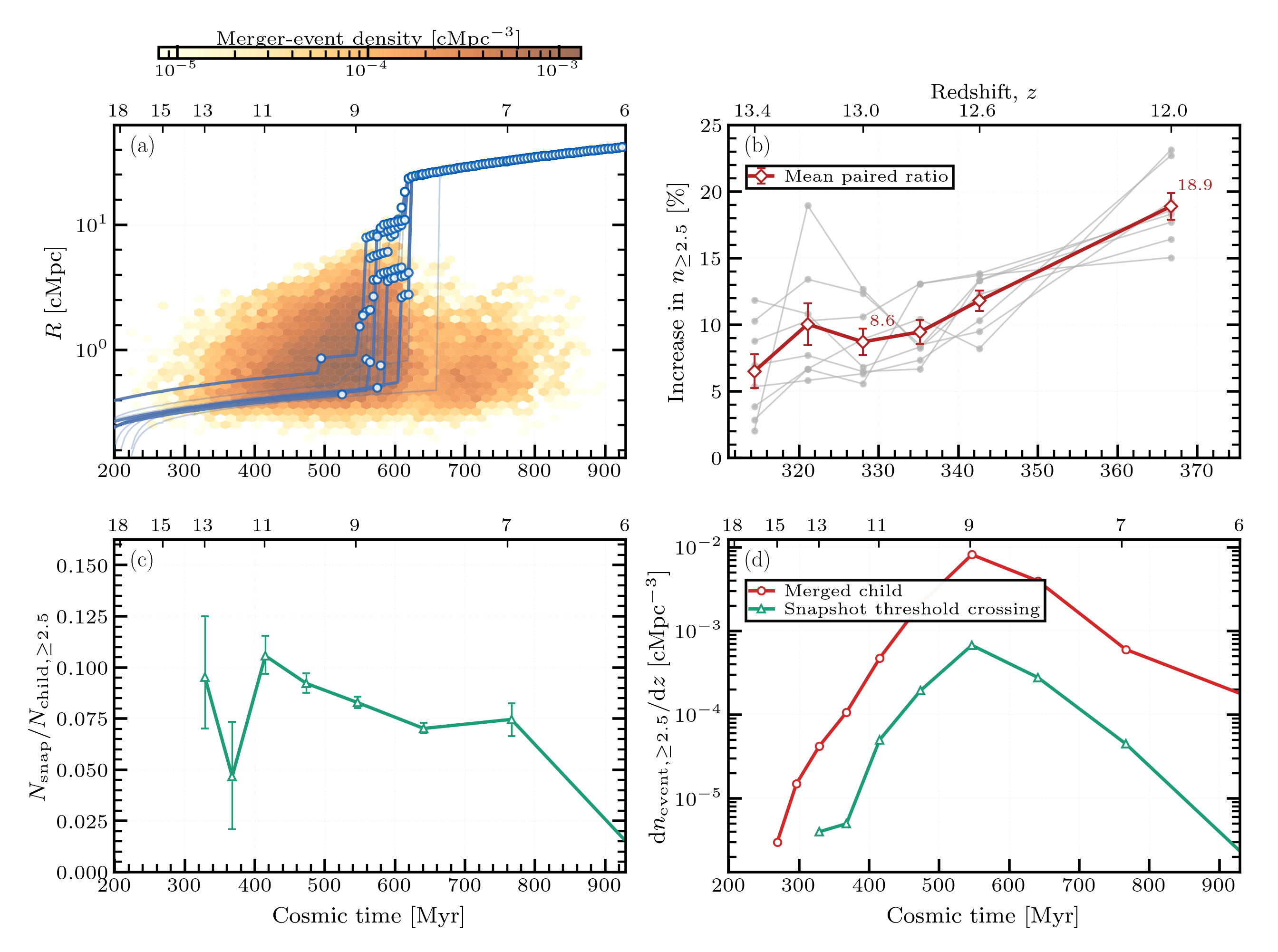}
    \caption{Spherical-merger results. Panels~(a), (c), and (d) use eight $L=50$~cMpc full-history realizations: panel~(a) shows representative connected-region tracks over merger-event density, panel~(c) gives the fraction of matched large-child events whose available parent snapshots are all below $R=2.5$~cMpc ($N_{\rm snap}/N_{\rm child,\ge2.5}$), and panel~(d) gives the densities of matched large-child events and this subset. The matching rule in \S\ref{sec:discussion_mergers} permits incomplete parent matches and compares different snapshot times, so the subset does not isolate merger-caused crossings. Panel~(b) instead shows the eight-realization $L=250$~cMpc correction to $n_{\ge2.5}$; thin curves are individual box ratios and symbols give their mean and standard error, while labels report the stacked-count ratio. Cosmic time is the primary coordinate and redshift is shown above.}
    \label{fig:merger_percolation_combined}
\end{figure*}

\subsection{Effect of spherical overlap and mergers}\label{sec:discussion_mergers}

The fiducial incidence counts non-merging source-group spheres. Figure~\ref{fig:merger_percolation_combined} uses an auxiliary spherical-merger calculation: panels~(a), (c), and (d) summarize eight $L=50$~cMpc full histories, while panel~(b) compares eight $L=250$~cMpc realizations evolved with and without mergers using the same random seeds and source populations.

The discrete jumps in Fig.~\ref{fig:merger_percolation_combined}(a) show how mergers redirect individual histories, and the event-density background shows that most merger activity involves smaller regions. Panel~(b) gives the headline large-volume comparison directly: the stacked $n_{\ge2.5}$ count increases from 743 to 807 ($8.6\%$) at $z=13$ and by $18.9\%$ at $z=12$. Over the same five-slice redshift window used for the headline incidence, the surface density rises from $8.10\times10^{-3}$ to $8.88\times10^{-3}$~arcmin$^{-2}$, an increase of $9.5\%$. The eight realization ratios give a mean enhancement of $8.7\pm1.0\%$ at $z=13$, where the quoted uncertainty is the standard error across realizations.

Panels~(c,d) compare child and parent radii at the event output, using the next child or previous parent output when necessary. Events require positive finite radii for the child and at least one parent. The subset $N_{\rm snap}$, with a large child but no matched parent above the threshold, comprises approximately $5$--$12\%$ of matched large-child events at $z\approx7$--12, declining toward $z\approx6$. Incomplete parent matches and growth between snapshots prevent interpreting this fraction as a merger-only promotion probability; panel~(b) measures the incidence change between the paired runs.

The merger prescription conserves the sum of the parent volumes, including overlapping contributions. Its spherical radii describe this volume prescription; they do not determine the connected path lengths relevant to Ly$\alpha$ transmission. Because the sources are placed randomly, this test does not include the additional, statistic-dependent effect of galaxy clustering.

\section{Conclusions}\label{sec:conclusions}

We have estimated the sky surface density $\Sigma_{\ge2.5}$ of non-merging source-group spheres with $R\ge2.5$~cMpc at $z\approx13$ using a UVLF-calibrated model. Our main results are:
\begin{enumerate}
\item \textbf{Witstok-sized bubbles can arise in the continuously active fiducial population.} In the \citet{Donnan2024PRIMER} UVLF model with $f_{\rm esc}=0.2$, $\log_{10}\xi_{\rm ion}=25.5$, $f_{\rm active}=1$, and $C=3$, we find $\Sigma_{\ge 2.5}\simeq 8.10\times 10^{-3}$ arcmin$^{-2}$ per $\Delta z=1$ at $z\approx 13$, or $\sim 0.32$ model spheres above the radius threshold in a $40$~arcmin$^2$ field per unit $\Delta z$. Lower active fractions, implemented through a larger latent reservoir that preserves the active UVLF, reduce both this incidence and the characteristic radius at fixed high-tail abundance. Persistent time-duty tests show that longer correlated on episodes strengthen the tail at fixed mean emissivity. The Bouwens Schechter comparison produces a further decrease of orders of magnitude. The inferred abundance is therefore conditional on both the source activity history and the adopted UVLF.

\item \textbf{The specific Witstok source may be atypical.} Its ionized environment may require unusually high effective $f_{\rm esc}\xi_{\rm ion}$, a recent brighter phase, or an AGN contribution. Population incidence alone cannot distinguish these interpretations.

\item \textbf{The spherical-merger prescription modestly increases threshold incidence.} In eight $L=250$~cMpc realizations evolved with both treatments, merging overlapping spheres increases the five-slice $\Sigma_{\ge2.5}$ by $9.5\%$.
\end{enumerate}

\noindent\textbf{Limitations and implications.}
The model holds on-state source luminosities fixed, places sources randomly, and represents unresolved sinks with an effective clumping factor. The latent $\phi_{\rm UVLF}/f$ population calibrates the emitting UVLF but is not independently constrained by stellar masses or halos; neither activity model follows burst-and-fading evolution. A halo-linked model with stochastic star formation would permit joint UVLF and stellar-mass-function tests. Source clustering \citep{Trapp2023,Naidu2020,Mason2025}, anisotropic radiative transfer, and peculiar velocities affect the connection between spherical radii and Ly$\alpha$ sight lines, so our incidence is not a strict lower bound on Ly$\alpha$ visibility. Predicting observed detections also requires emitter occupation and survey selection. Deeper \textit{JWST} imaging and spectroscopy can refine the UVLF and constrain source histories \citep{Donnan2024PRIMER,Tang2024,Witstok2025Nature,Mason2025}.

\begin{acknowledgments}
We thank Tobias Marriage and Stephen McCandliss for helpful discussions and comments that improved this work. We thank the anonymous referee for constructive suggestions that improved the manuscript. MS acknowledges support for this work under NASA grant 80NSSC22K1294. This work made use of ChatGPT and Claude for code review and debugging, though scientific content and interpretations are the responsibility of the authors. The simulation data and analysis scripts used in this work are available from the corresponding author upon reasonable request.
\end{acknowledgments}

\FloatBarrier
\appendix
\renewcommand{\theHequation}{\Alph{section}.\arabic{equation}}
\section{Analytic models for the bubble size distribution}\label{app:analytic_bsd}

We compare three mappings from the UVLF to the BSD of spherical regions before percolation. Method 1 uses an effective emission duration, Method 2 follows source cohorts and approximates their grouping, and Method 3 preserves luminosity rank as the UVLF evolves. We write the UVLF per unit magnitude as $\phi(M_{\rm UV},z)$.

\subsection{Method 1: Instantaneous UVLF mapping}

In the photon-counting approximation, a source with effective emission duration $t_{\rm eff}$ produces comoving volume
\begin{equation}
V\simeq\frac{\dot N_{\rm ion}(z)}{\bar n_{\rm H,0}}\,t_{\rm eff}.
\label{eq:m1-volume}
\end{equation}
At fixed $t_{\rm eff}$ and emissivity conversion, $R\propto10^{-0.4M_{\rm UV}/3}$ gives
\begin{equation}
\frac{\mathrm{d}n}{\mathrm{d}R}=\phi(M_{\rm UV}(R),z)\frac{7.5}{R\ln10}.
\label{eq:m1-bsd}
\end{equation}
Figure~\ref{fig:hist_z13} uses an illustrative $t_{\rm eff}=50$~Myr. More generally, $t_{\rm eff}=\int\dot N_{\rm ion}(t')\,\mathrm{d}t'/\dot N_{\rm ion}(z)$ weights the emission history and can differ from the elapsed source age.

\subsection{Method 2: Age-cohort synthesis with multiplicity correction}\label{app:method2}

\subsubsection{Seed histories}
At fixed magnitude, the positive growth of the UVLF defines a birth measure per unit redshift,
\begin{equation}
B(z_b,M_{\rm UV})=\max[-\partial_{z_b}\phi(M_{\rm UV},z_b),0],
\label{eq:birth-rate}
\end{equation}
The initial population at $z_{\max}=25$ starts with zero ionized volume.

For birth time $t_b=t(z_b)$ and observation time $t_o=t(z_{\rm obs})$, define the recombination survival kernel and effective emitting duration:
\begin{equation}
\mathcal S(t,t_o)=\exp\!\left[-\int_t^{t_o}\frac{\mathrm{d}t'}{t_{\rm rec}(t')}\right],
\qquad
\tau_{\rm eff}(z_b,z_{\rm obs})=\int_{t_b}^{t_o}\mathcal S(t,t_o)\,\mathrm{d}t,
\label{eq:tau-eff}
\end{equation}
A continuously emitting cohort has volume
\begin{equation}
V(z_b,M_{\rm UV})=\frac{\dot N_{\rm ion}(M_{\rm UV})}{\bar n_{\rm H,0}}\,\tau_{\rm eff}(z_b,z_{\rm obs}),
\qquad R(z_b,M_{\rm UV})=\left[\frac{3V(z_b,M_{\rm UV})}{4\pi}\right]^{1/3}.
\label{eq:v-of-zb-M}
\end{equation}
The seed BSD, $\psi_{\rm seed}=\mathrm{d}n/\mathrm{d}R$, is
\begin{equation}
\psi_{\rm seed}(R)=\psi_{\rm init}(R)+\int_{z_{\rm obs}}^{z_{\max}}\!\mathrm{d}z_b\int\!\mathrm{d}M_{\rm UV}\,B(z_b,M_{\rm UV})\,\delta_{\rm D}[R-R(z_b,M_{\rm UV})],
\label{eq:psi1}
\end{equation}
with the initial-cohort contribution
\begin{equation}
\psi_{\rm init}(R)=\int\!\mathrm{d}M_{\rm UV}\,\phi(M_{\rm UV},z_{\max})\,\delta_{\rm D}[R-R(z_{\max},M_{\rm UV})].
\label{eq:psi-init}
\end{equation}

For luminosity-dependent efficiencies, we use $\dot N_{\rm ion}(M_{\rm UV},z)=f_{\rm esc}(M_{\rm UV},z)\xi_{\rm ion}(M_{\rm UV},z)L_{\rm UV}(M_{\rm UV})$ inside the survival integral, $V=\bar n_{\rm H,0}^{-1}\int_{t_b}^{t_o}\dot N_{\rm ion}(M_{\rm UV},t)\mathcal S(t,t_o)\,\mathrm{d}t$. The seed mapping sums all monotonic magnitude--radius branches, including both branches of the \citet{Anderson2017} prescription, and the companion moments below use these same volumes.

\subsubsection{Companion multiplicity}
For a seed of volume $V_1=4\pi R_1^3/3$, we approximate the number of additional sources, $K=N-1$, by
\begin{equation}
P(K=k\mid R_1)=\frac{\mu(R_1)^k e^{-\mu(R_1)}}{k!},
\qquad \mu(R_1)=V_1\lambda_n,
\label{eq:poisson}
\end{equation}
using the effective companion density
\begin{equation}
\lambda_n=\frac{1}{z_{\max}-z_{\rm obs}}\int_{z_{\rm obs}}^{z_{\max}}\!\mathrm{d}z\int\!\mathrm{d}M_{\rm UV}\,\phi(M_{\rm UV},z).
\label{eq:companion-density}
\end{equation}
The corresponding individual-volume moments are
\begin{equation}
\langle V_c^j\rangle=
\frac{\displaystyle\int_{z_{\rm obs}}^{z_{\max}}\!\mathrm{d}z\int\!\mathrm{d}M_{\rm UV}\,\phi(M_{\rm UV},z)V(z,M_{\rm UV})^j}
{\displaystyle\int_{z_{\rm obs}}^{z_{\max}}\!\mathrm{d}z\int\!\mathrm{d}M_{\rm UV}\,\phi(M_{\rm UV},z)},
\qquad j=1,2,
\label{eq:companion-moments}
\end{equation}
where $V(z,M_{\rm UV})$ is evaluated at $z_{\rm obs}$ with the same survival kernel as the seeds. Write $\bar V_c=\langle V_c\rangle$ and $\sigma_{V,c}^2=\langle V_c^2\rangle-\bar V_c^2$. For $k\ge1$, the summed companion volume $\Delta V_k$ is approximated by a gamma distribution $p_k$ with shape $a_k=k\bar V_c^2/\sigma_{V,c}^2$ and scale $\theta=\sigma_{V,c}^2/\bar V_c$. Zero variance gives deterministic $\Delta V_k=k\bar V_c$, and $k=0$ adds no volume.

The radius kernel and multiplicity components are
\begin{equation}
\mathcal K_{k+1}(R\mid R_1)=\int\!\mathrm{d}\Delta V_k\,p_k(\Delta V_k)\,
\delta_{\rm D}\!\left[R-\left(\frac{3(V_1+\Delta V_k)}{4\pi}\right)^{1/3}\right],
\label{eq:multiplicity_kernel}
\end{equation}
\begin{equation}
\psi_N(R)=\int\!\mathrm{d}R_1\,\psi_{\rm seed}(R_1)P(K=N-1\mid R_1)\mathcal K_N(R\mid R_1),
\qquad \psi(R)=\sum_{N\ge1}\psi_N(R).
\label{eq:psiN}
\end{equation}
The $N=1$ term includes the no-companion probability. This approximation leaves seeds undepleted, omits recursive attachment and clustering, and does not partition sources uniquely into groups. Alternative companion-density and volume-weighting prescriptions change the fiducial analytic tail by up to $19\%$.

\begin{figure*}[!tbp]
    \centering
    \includegraphics[width=0.96\textwidth]{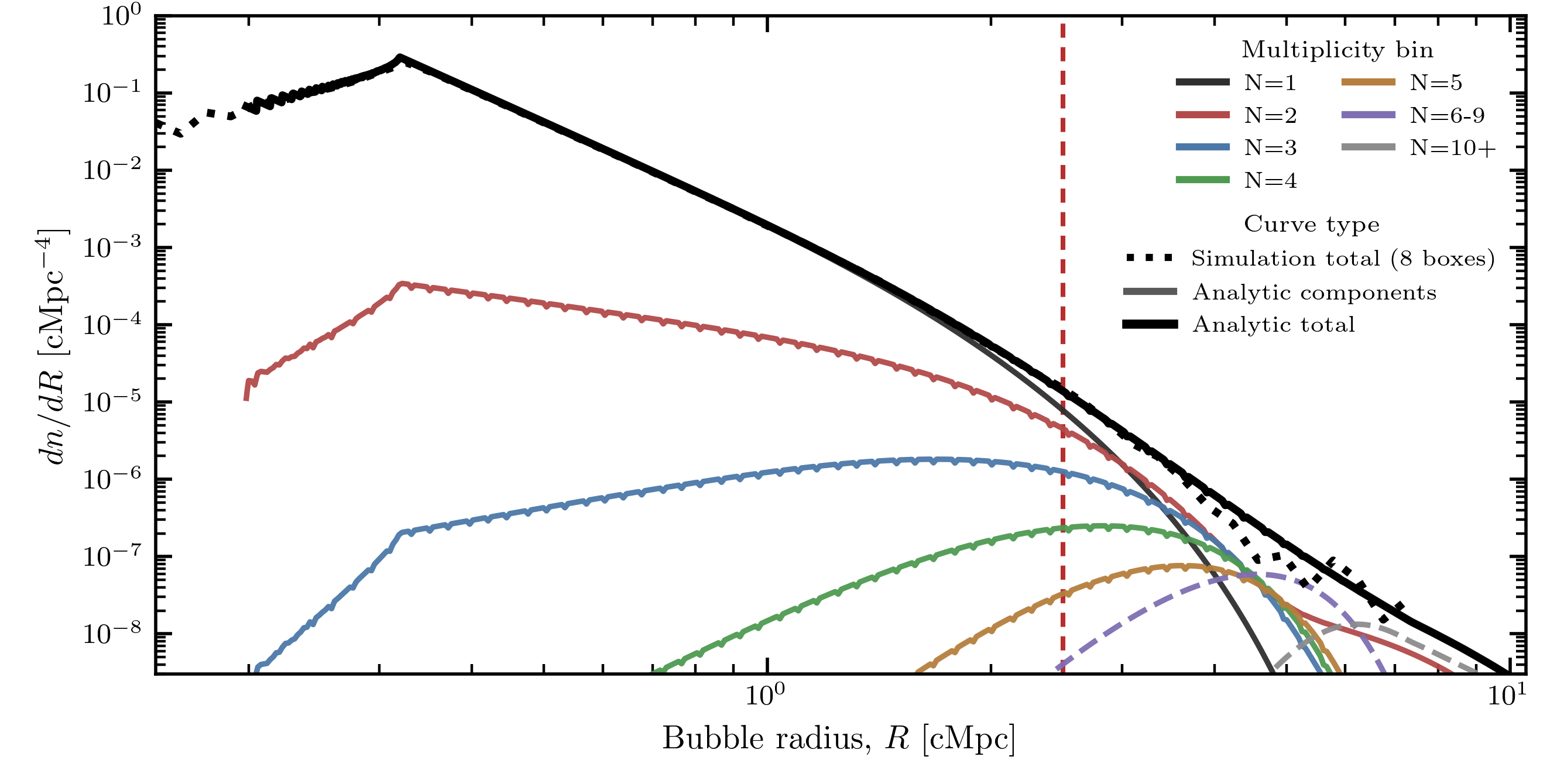}
    \caption{Method-2 multiplicity components at \texorpdfstring{$z\simeq13$}{z~13}. The analytic label $N$ denotes one seed plus $N-1$ Poisson companions in Eq.~\eqref{eq:psiN}. Colored curves show the gamma companion-volume model, the solid black curve is their sum, and the dotted black curve is the total BSD from eight fiducial $L=250$~cMpc simulations. The analytic labels do not identify measured simulation memberships. The red line marks $R=2.5$~cMpc.}
    \label{fig:multiplicity_components}
\end{figure*}

\subsubsection{Intermittent source activity}\label{app:active_bsd_correction}
For activity $I(t)=1$ when on and 0 when off, with fixed on-state luminosity, define
\begin{equation}
W=\frac{\displaystyle\int_{t_b}^{t_o}I(t)\mathcal S(t,t_o)\,\mathrm{d}t}
{\displaystyle\int_{t_b}^{t_o}\mathcal S(t,t_o)\,\mathrm{d}t},
\qquad V_f=W V_{\rm cont},\qquad R_f=W^{1/3}R_{\rm cont},
\label{eq:weighted_activity}
\end{equation}
where $V_{\rm cont}$ and $R_{\rm cont}$ are the continuously emitting volume and radius in Eq.~\eqref{eq:v-of-zb-M}. We use $f$ for either $f_{\rm active}$ or $f_{\rm time}$ below. The two activity prescriptions determine the mean $\mu_W=f$ and variance $\sigma_W^2$ for each cohort. We approximate their distribution by $W\sim\mathrm{Beta}(a_W,b_W)$, with
\begin{equation}
a_W=\mu_W\left[\frac{\mu_W(1-\mu_W)}{\sigma_W^2}-1\right],
\qquad b_W=(1-\mu_W)\left[\frac{\mu_W(1-\mu_W)}{\sigma_W^2}-1\right].
\label{eq:beta_activity}
\end{equation}
Negligible variance gives $W=\mu_W$; maximal variance $\mu_W(1-\mu_W)$ gives Bernoulli masses at zero and one. Between these limits, the continuous beta approximation omits the discrete all-off and all-on probabilities of a finite history.

Writing $\psi_{\rm cont}(R\mid z_b)$ for the continuous seed contribution per unit birth redshift, the intermittent BSD at $R>0$ is
\begin{equation}
\begin{aligned}
\psi_f(R)&=\psi_{\rm init,f}(R)+\frac{1}{f}\int_{z_{\rm obs}}^{z_{\max}}\!\mathrm{d}z_b
\int_{(0,1]}\!W^{-1/3}\psi_{\rm cont}\!\left(\frac{R}{W^{1/3}}\middle|z_b\right)\mathrm{d}P(W\mid z_b).
\end{aligned}
\label{eq:activity_bsd}
\end{equation}
The measure $\mathrm{d}P$ includes continuous and discrete positive activity; all-off histories contribute at $R=0$. The initial cohort receives the same transformation. The factor $1/f$ preserves the emitting UVLF through the larger latent population; companion calculations likewise use $\lambda_n/f$ and activity-remapped volume moments.

\noindent\textit{Active-fraction model.}
With $N_{\rm obs}$ emitting galaxies among $N_{\rm lat}$ total galaxies in a magnitude bin, independent redraws give each member activation probability $N_{\rm obs}/N_{\rm lat}\simeq f_{\rm active}$. Selection without replacement correlates different members but preserves this marginal history. For 5~Myr lookback bins, with a shortened final bin and midpoint survival values $\mathcal S_j$,
\begin{equation}
W=\sum_jw_j I_j,\qquad w_j=\frac{\mathcal S_j\Delta t_j}{\sum_k\mathcal S_k\Delta t_k},
\qquad \langle W\rangle=f_{\rm active},\qquad
\sigma_W^2=f_{\rm active}(1-f_{\rm active})\sum_jw_j^2.
\label{eq:active_moments}
\end{equation}

\noindent\textit{Time-duty model.}
The on and off transition rates, $\lambda_+=1/\langle t_{\rm off}\rangle$ and $\lambda_-=1/\langle t_{\rm on}\rangle$, give
\begin{equation}
f_{\rm time}=\frac{\lambda_+}{\lambda_++\lambda_-},\qquad
\tau_c=\frac{1}{\lambda_++\lambda_-}=\langle t_{\rm on}\rangle(1-f_{\rm time}),
\label{eq:activity_correlation}
\end{equation}
where $\tau_c$ is the activity correlation time. The stationary covariance is $\mathrm{Cov}[I(t),I(t+\Delta t)]=f_{\rm time}(1-f_{\rm time})e^{-|\Delta t|/\tau_c}$. Thus $\langle W\rangle=f_{\rm time}$ and
\begin{equation}
\sigma_W^2=f_{\rm time}(1-f_{\rm time})
\frac{\displaystyle\int_{t_b}^{t_o}\!\mathrm{d}t\int_{t_b}^{t_o}\!\mathrm{d}t'\,
\mathcal S(t,t_o)\mathcal S(t',t_o)e^{-|t-t'|/\tau_c}}
{\displaystyle\left[\int_{t_b}^{t_o}\mathcal S(t,t_o)\,\mathrm{d}t\right]^2}.
\label{eq:time_moments}
\end{equation}
Longer correlations broaden the distribution of cumulative activity and can strengthen the large-radius tail at fixed mean emissivity. Both analytic activity predictions use the UVLF and growth law without fitting to the simulated BSD. 

\subsection{Method 3: Abundance matching}\label{app:method3}

Method 3 preserves luminosity rank, labeled by cumulative comoving density
\begin{equation}
n_{\rm cum}=n(<M_{\rm UV},z)=\int_{-\infty}^{M_{\rm UV}}\phi(M_{\rm UV}',z)\,\mathrm{d}M_{\rm UV}'.
\end{equation}
Each rank evolves according to
\begin{equation}
\frac{\mathrm{d}V}{\mathrm{d}t}=\frac{\dot N_{\rm ion}[M_{\rm UV}(n_{\rm cum},t)]}{\bar n_{\rm H,0}}-\frac{V}{t_{\rm rec}(z)}.
\label{eq:m3-ode}
\end{equation}
Ranks already present at $z_{\max}$ start with zero volume; additional ranks begin emitting as $n(<M_{\rm lim},z)$ grows past them. On a monotonic radius--rank branch,
\begin{equation}
\frac{\mathrm{d}n}{\mathrm{d}R}=\left|\frac{\mathrm{d}R}{\mathrm{d}n_{\rm cum}}\right|^{-1}
=\frac{n_{\rm cum}}{R}\left|\frac{\mathrm{d}\ln n_{\rm cum}}{\mathrm{d}\ln R}\right|,
\label{eq:m3-bsd}
\end{equation}
with the logarithmic derivative evaluated numerically. Rank preservation evolves the assigned luminosity with the UVLF and omits stochastic reshuffling, bursts, and progenitor mergers. No companion correction is applied.

The three methods agree to order unity over $R\sim0.3$--3~cMpc. Method 2 gives $n_{\ge2.5}=6.25\times10^{-6}$~cMpc$^{-3}$, compared with the simulated $5.94\times10^{-6}$~cMpc$^{-3}$. Figure~\ref{fig:multiplicity_components} shows the increasing contribution of multiple sources to its large-radius tail.

\FloatBarrier
\clearpage
\section{Full parameter study}\label{app:full_parameter_suite}
Table~\ref{tab:sim_bubbles_full} reports all 56 parameter combinations, including the luminosity-dependent prescriptions and both activity models, each evaluated in eight $L=250$~cMpc boxes.

\startlongtable
\begin{deluxetable*}{ccccccccccccc}
\tabletypesize{\tiny}
\renewcommand{\arraystretch}{0.88}
\setlength{\tabcolsep}{2.5pt}
\tablecaption{Simulation results and analytic predictions for all models\label{tab:sim_bubbles_full}}
\tablehead{
\colhead{UVLF / mode} & \colhead{$f_{\mathrm{esc}}$} & \colhead{$\log \xi_{\mathrm{ion}}$} & \colhead{$f$} & \colhead{$C$} & \colhead{$M_{\mathrm{lim}}$} & \colhead{$x_{\mathrm{HII}}^{\rm phot}$} & \colhead{$x_{\mathrm{HII}}^{\rm bub}$} & \colhead{$n_{\ge2.5}^{\rm sim}$} & \colhead{$n_{\ge2.5}^{\rm ana}$} & \colhead{$\Sigma_{\ge2.5}^{\rm mid}$} & \colhead{$\Sigma_{\ge2.5}^{\rm 5s}$} & \colhead{$\Sigma_{\ge2.5}^{\rm ana}$} \\
\colhead{} & \colhead{} & \colhead{} & \colhead{} & \colhead{} & \colhead{} & \colhead{terminal} & \colhead{terminal} & \colhead{[cMpc$^{-3}$]} & \colhead{[cMpc$^{-3}$]} & \colhead{[arcmin$^{-2}$]} & \colhead{[arcmin$^{-2}$ per $\Delta z\!=\!1$]} & \colhead{[arcmin$^{-2}$]}
}
\startdata
B21 & 0.2 & 25.5 & 0.1 & 3 & -13.0 & 1.00 & 0.994 & $0^{+1.5}_{-0}\times10^{-8}$ & $5.82\times10^{-9}$ & $0$ & $0^{+2}_{-0}\times10^{-5}$ & $9.03\times10^{-6}$ \\
B21 & 0.2 & 25.5 & 0.2 & 3 & -13.0 & 1.00 & 0.992 & $0^{+1.5}_{-0}\times10^{-8}$ & $1.07\times10^{-8}$ & $0$ & $0^{+2}_{-0}\times10^{-5}$ & $1.68\times10^{-5}$ \\
B21 & 0.2 & 25.5 & 0.3 & 3 & -13.0 & 1.00 & 0.992 & $8.00^{+18}_{-8}\times10^{-9}$ & $1.69\times10^{-8}$ & $1.06\times10^{-5}$ & $1.29^{+0.86}_{-0.86}\times10^{-5}$ & $2.65\times10^{-5}$ \\
B21 & 0.2 & 25.5 & 0.4 & 3 & -13.0 & 1.00 & 0.992 & $0^{+1.5}_{-0}\times10^{-8}$ & $2.41\times10^{-8}$ & $0$ & $6.54^{+6.4}_{-4.3}\times10^{-6}$ & $3.78\times10^{-5}$ \\
B21 & 0.2 & 25.5 & 0.5 & 3 & -13.0 & 1.00 & 0.992 & $2.40^{+2.3}_{-1.3}\times10^{-8}$ & $3.20\times10^{-8}$ & $3.18\times10^{-5}$ & $2.77^{+1.3}_{-1.1}\times10^{-5}$ & $5.01\times10^{-5}$ \\
B21 & 0.2 & 25.5 & 0.6 & 3 & -13.0 & 1.00 & 0.991 & $3.20^{+2.5}_{-1.5}\times10^{-8}$ & $4.05\times10^{-8}$ & $4.25\times10^{-5}$ & $3.42^{+1.3}_{-1.3}\times10^{-5}$ & $6.32\times10^{-5}$ \\
B21 & 0.2 & 25.5 & 0.7 & 3 & -13.0 & 1.00 & 0.992 & $3.20^{+2.5}_{-1.5}\times10^{-8}$ & $4.93\times10^{-8}$ & $4.25\times10^{-5}$ & $3.21^{+1.3}_{-1.3}\times10^{-5}$ & $7.69\times10^{-5}$ \\
B21 & 0.2 & 25.5 & 0.8 & 3 & -13.0 & 1.00 & 0.991 & $4.00^{+2.7}_{-1.7}\times10^{-8}$ & $5.84\times10^{-8}$ & $5.31\times10^{-5}$ & $4.08^{+1.2}_{-1.1}\times10^{-5}$ & $9.09\times10^{-5}$ \\
B21 & 0.2 & 25.5 & 0.9 & 3 & -13.0 & 1.00 & 0.990 & $3.20^{+2.5}_{-1.5}\times10^{-8}$ & $6.77\times10^{-8}$ & $4.25\times10^{-5}$ & $3.85^{+1.3}_{-1.3}\times10^{-5}$ & $1.05\times10^{-4}$ \\
B21 & 0.2 & 25.5 & 1.0 & 3 & -13.0 & 1.00 & 0.991 & $4.80^{+2.9}_{-1.9}\times10^{-8}$ & $7.72\times10^{-8}$ & $6.37\times10^{-5}$ & $6.83^{+1.7}_{-1.7}\times10^{-5}$ & $1.20\times10^{-4}$ \\
B21-T(10) & 0.2 & 25.5 & 0.1 & 3 & -13.0 & 1.00 & 0.993 & $8.00^{+18}_{-8}\times10^{-9}$ & $2.1\times10^{-8}$ & $1.06\times10^{-5}$ & $1.52^{+0.82}_{-0.65}\times10^{-5}$ & $3.2\times10^{-5}$ \\
B21-T(2.5) & 0.2 & 25.5 & 0.2 & 3 & -13.0 & 1.00 & 0.992 & $0^{+1.5}_{-0}\times10^{-8}$ & $8.57\times10^{-9}$ & $0$ & $0^{+2}_{-0}\times10^{-5}$ & $1.34\times10^{-5}$ \\
B21-T(10) & 0.2 & 25.5 & 0.2 & 3 & -13.0 & 1.00 & 0.992 & $8.00^{+18}_{-8}\times10^{-9}$ & $2.41\times10^{-8}$ & $1.06\times10^{-5}$ & $1.72^{+1.3}_{-1.3}\times10^{-5}$ & $3.69\times10^{-5}$ \\
B21-T(40) & 0.2 & 25.5 & 0.2 & 3 & -13.0 & 1.00 & 0.992 & $3.20^{+2.5}_{-1.6}\times10^{-8}$ & $5.06\times10^{-8}$ & $4.25\times10^{-5}$ & $4.26^{+2.3}_{-2.3}\times10^{-5}$ & $7.76\times10^{-5}$ \\
B21-T(10) & 0.2 & 25.5 & 0.4 & 3 & -13.0 & 1.00 & 0.992 & $3.20^{+2.5}_{-2.4}\times10^{-8}$ & $3.20\times10^{-8}$ & $4.25\times10^{-5}$ & $3.64^{+2.6}_{-2.8}\times10^{-5}$ & $4.95\times10^{-5}$ \\
B21-T(10) & 0.2 & 25.5 & 0.5 & 3 & -13.0 & 1.00 & 0.992 & $2.40^{+2.3}_{-1.6}\times10^{-8}$ & $3.7\times10^{-8}$ & $3.18\times10^{-5}$ & $3.24^{+1.3}_{-1.3}\times10^{-5}$ & $5.73\times10^{-5}$ \\
B21-T(10) & 0.2 & 25.5 & 0.6 & 3 & -13.0 & 1.00 & 0.991 & $3.20^{+2.5}_{-1.6}\times10^{-8}$ & $4.29\times10^{-8}$ & $4.25\times10^{-5}$ & $4.51^{+2.1}_{-1.9}\times10^{-5}$ & $6.65\times10^{-5}$ \\
B21-T(10) & 0.2 & 25.5 & 0.8 & 3 & -13.0 & 1.00 & 0.992 & $7.20^{+3.3}_{-2.4}\times10^{-8}$ & $5.76\times10^{-8}$ & $9.55\times10^{-5}$ & $8.33^{+2.5}_{-2.5}\times10^{-5}$ & $8.92\times10^{-5}$ \\
D24 & 0.2 & 25.5 & 0.1 & 3 & -13.0 & 1.00 & 1.00 & $9.76^{+1.9}_{-1.9}\times10^{-7}$ & $1.40\times10^{-6}$ & $1.29\times10^{-3}$ & $1.36^{+0.23}_{-0.23}\times10^{-3}$ & $1.90\times10^{-3}$ \\
D24 & 0.2 & 25.5 & 0.2 & 3 & -13.0 & 1.00 & 1.00 & $1.45^{+0.14}_{-0.13}\times10^{-6}$ & $1.98\times10^{-6}$ & $1.92\times10^{-3}$ & $1.97^{+0.17}_{-0.17}\times10^{-3}$ & $2.70\times10^{-3}$ \\
D24 & 0.2 & 25.5 & 0.3 & 3 & -13.0 & 1.00 & 1.00 & $1.95^{+0.11}_{-0.11}\times10^{-6}$ & $2.56\times10^{-6}$ & $2.59\times10^{-3}$ & $2.69^{+0.12}_{-0.12}\times10^{-3}$ & $3.49\times10^{-3}$ \\
D24 & 0.2 & 25.5 & 0.4 & 3 & -13.0 & 1.00 & 1.00 & $2.50^{+0.14}_{-0.14}\times10^{-6}$ & $3.12\times10^{-6}$ & $3.32\times10^{-3}$ & $3.35^{+0.15}_{-0.15}\times10^{-3}$ & $4.26\times10^{-3}$ \\
D24 & 0.2 & 25.5 & 0.5 & 3 & -13.0 & 1.00 & 1.00 & $2.92^{+0.18}_{-0.18}\times10^{-6}$ & $3.68\times10^{-6}$ & $3.87\times10^{-3}$ & $4.07^{+0.22}_{-0.22}\times10^{-3}$ & $5.02\times10^{-3}$ \\
D24 & 0.2 & 25.5 & 0.6 & 3 & -13.0 & 1.00 & 1.00 & $3.66^{+0.22}_{-0.22}\times10^{-6}$ & $4.22\times10^{-6}$ & $4.85\times10^{-3}$ & $5.00^{+0.28}_{-0.28}\times10^{-3}$ & $5.76\times10^{-3}$ \\
D24 & 0.2 & 25.5 & 0.7 & 3 & -13.0 & 1.00 & 1.00 & $4.30^{+0.18}_{-0.18}\times10^{-6}$ & $4.74\times10^{-6}$ & $5.71\times10^{-3}$ & $5.83^{+0.23}_{-0.23}\times10^{-3}$ & $6.48\times10^{-3}$ \\
D24 & 0.2 & 25.5 & 0.8 & 3 & -13.0 & 1.00 & 1.00 & $4.87^{+0.22}_{-0.22}\times10^{-6}$ & $5.26\times10^{-6}$ & $6.46\times10^{-3}$ & $6.70^{+0.26}_{-0.26}\times10^{-3}$ & $7.19\times10^{-3}$ \\
D24 & 0.2 & 25.5 & 0.9 & 3 & -13.0 & 1.00 & 1.00 & $5.41^{+0.3}_{-0.3}\times10^{-6}$ & $5.76\times10^{-6}$ & $7.18\times10^{-3}$ & $7.53^{+0.29}_{-0.29}\times10^{-3}$ & $7.88\times10^{-3}$ \\
D24$^{\dagger}$ & 0.2 & 25.5 & 1.0 & 3 & -13.0 & 1.00 & 1.00 & $5.94^{+0.2}_{-0.2}\times10^{-6}$ & $6.25\times10^{-6}$ & $7.89\times10^{-3}$ & $8.10^{+0.25}_{-0.24}\times10^{-3}$ & $8.55\times10^{-3}$ \\
D24-T(10) & 0.2 & 25.5 & 0.1 & 3 & -13.0 & 1.00 & 1.00 & $1.64^{+0.15}_{-0.15}\times10^{-6}$ & $2.27\times10^{-6}$ & $2.18\times10^{-3}$ & $2.26^{+0.14}_{-0.14}\times10^{-3}$ & $3.08\times10^{-3}$ \\
D24-T(2.5) & 0.2 & 25.5 & 0.2 & 3 & -13.0 & 1.00 & 1.00 & $1.32^{+0.11}_{-0.11}\times10^{-6}$ & $1.89\times10^{-6}$ & $1.75\times10^{-3}$ & $1.77^{+0.15}_{-0.15}\times10^{-3}$ & $2.57\times10^{-3}$ \\
D24-T(10) & 0.2 & 25.5 & 0.2 & 3 & -13.0 & 1.00 & 1.00 & $2.08^{+0.12}_{-0.12}\times10^{-6}$ & $2.65\times10^{-6}$ & $2.76\times10^{-3}$ & $2.81^{+0.11}_{-0.11}\times10^{-3}$ & $3.59\times10^{-3}$ \\
D24-T(40) & 0.2 & 25.5 & 0.2 & 3 & -13.0 & 1.00 & 1.00 & $3.74^{+0.21}_{-0.21}\times10^{-6}$ & $4.18\times10^{-6}$ & $4.96\times10^{-3}$ & $5.00^{+0.22}_{-0.22}\times10^{-3}$ & $5.69\times10^{-3}$ \\
D24-T(10) & 0.2 & 25.5 & 0.4 & 3 & -13.0 & 1.00 & 1.00 & $2.99^{+0.11}_{-0.11}\times10^{-6}$ & $3.44\times10^{-6}$ & $3.97\times10^{-3}$ & $4.00^{+0.17}_{-0.17}\times10^{-3}$ & $4.69\times10^{-3}$ \\
D24-T(10) & 0.2 & 25.5 & 0.5 & 3 & -13.0 & 1.00 & 1.00 & $3.24^{+0.18}_{-0.18}\times10^{-6}$ & $3.86\times10^{-6}$ & $4.3\times10^{-3}$ & $4.39^{+0.23}_{-0.23}\times10^{-3}$ & $5.26\times10^{-3}$ \\
D24-T(10) & 0.2 & 25.5 & 0.6 & 3 & -13.0 & 1.00 & 1.00 & $3.98^{+0.24}_{-0.24}\times10^{-6}$ & $4.30\times10^{-6}$ & $5.29\times10^{-3}$ & $5.37^{+0.28}_{-0.27}\times10^{-3}$ & $5.87\times10^{-3}$ \\
D24-T(10) & 0.2 & 25.5 & 0.8 & 3 & -13.0 & 1.00 & 1.00 & $5.13^{+0.24}_{-0.25}\times10^{-6}$ & $5.24\times10^{-6}$ & $6.80\times10^{-3}$ & $6.95^{+0.3}_{-0.3}\times10^{-3}$ & $7.15\times10^{-3}$ \\
D24 & 0.2 & 25.5 & 1.0 & 3 & -12.0 & 1.00 & 1.00 & $6.00^{+0.18}_{-0.18}\times10^{-6}$ & $6.32\times10^{-6}$ & $7.96\times10^{-3}$ & $8.20^{+0.25}_{-0.26}\times10^{-3}$ & $8.67\times10^{-3}$ \\
D24 & 0.2 & 25.5 & 1.0 & 3 & -12.5 & 1.00 & 1.00 & $6.00^{+0.2}_{-0.2}\times10^{-6}$ & $6.27\times10^{-6}$ & $7.96\times10^{-3}$ & $8.14^{+0.23}_{-0.23}\times10^{-3}$ & $8.60\times10^{-3}$ \\
D24 & 0.2 & 25.5 & 1.0 & 3 & -13.5 & 1.00 & 1.00 & $5.98^{+0.22}_{-0.22}\times10^{-6}$ & $6.22\times10^{-6}$ & $7.94\times10^{-3}$ & $8.18^{+0.27}_{-0.27}\times10^{-3}$ & $8.51\times10^{-3}$ \\
D24 & 0.2 & 25.5 & 1.0 & 3 & -14.0 & 1.00 & 0.999 & $5.90^{+0.19}_{-0.18}\times10^{-6}$ & $6.19\times10^{-6}$ & $7.83\times10^{-3}$ & $8.11^{+0.25}_{-0.25}\times10^{-3}$ & $8.48\times10^{-3}$ \\
D24 & 0.2 & 25.5 & 1.0 & 3 & -14.5 & 1.00 & 0.998 & $6.00^{+0.21}_{-0.21}\times10^{-6}$ & $6.17\times10^{-6}$ & $7.96\times10^{-3}$ & $8.16^{+0.26}_{-0.26}\times10^{-3}$ & $8.45\times10^{-3}$ \\
D24 & 0.2 & 25.5 & 1.0 & 1 & -13.0 & 1.00 & 1.00 & $1.22^{+0.0096}_{-0.0096}\times10^{-5}$ & $1.24\times10^{-5}$ & $1.61\times10^{-2}$ & $1.66^{+0.011}_{-0.011}\times10^{-2}$ & $1.70\times10^{-2}$ \\
D24 & 0.2 & 25.5 & 1.0 & 20 & -13.0 & 1.00 & 0.951 & $2.96^{+0.4}_{-0.4}\times10^{-7}$ & $4.24\times10^{-7}$ & $3.93\times10^{-4}$ & $4.14^{+0.49}_{-0.49}\times10^{-4}$ & $5.81\times10^{-4}$ \\
D24 & 0.2 & 25.5 & 1.0 & MD14 & -13.0 & 1.00 & 1.00 & $1.02^{+0.0088}_{-0.008}\times10^{-5}$ & $1.04\times10^{-5}$ & $1.35\times10^{-2}$ & $1.39^{+0.012}_{-0.013}\times10^{-2}$ & $1.42\times10^{-2}$ \\
D24 & 0.1 & 25.5 & 1.0 & 3 & -13.0 & 1.00 & 0.980 & $1.39^{+0.11}_{-0.11}\times10^{-6}$ & $1.67\times10^{-6}$ & $1.85\times10^{-3}$ & $1.90^{+0.13}_{-0.13}\times10^{-3}$ & $2.28\times10^{-3}$ \\
D24 & 0.3 & 25.5 & 1.0 & 3 & -13.0 & 1.00 & 1.00 & $1.29^{+0.014}_{-0.014}\times10^{-5}$ & $1.31\times10^{-5}$ & $1.71\times10^{-2}$ & $1.76^{+0.013}_{-0.013}\times10^{-2}$ & $1.80\times10^{-2}$ \\
D24 & 0.4 & 25.5 & 1.0 & 3 & -13.0 & 1.00 & 1.00 & $2.15^{+0.019}_{-0.019}\times10^{-5}$ & $2.17\times10^{-5}$ & $2.85\times10^{-2}$ & $2.93^{+0.013}_{-0.013}\times10^{-2}$ & $2.99\times10^{-2}$ \\
D24 & 0.2 & 24.5 & 1.0 & 3 & -13.0 & 0.542 & 0.454 & $9.60^{+3.6}_{-3.2}\times10^{-8}$ & $7.49\times10^{-8}$ & $1.27\times10^{-4}$ & $1.25^{+0.45}_{-0.45}\times10^{-4}$ & $1.01\times10^{-4}$ \\
D24 & 0.2 & 25.0 & 1.0 & 3 & -13.0 & 1.00 & 0.891 & $4.64^{+0.4}_{-0.4}\times10^{-7}$ & $6.87\times10^{-7}$ & $6.16\times10^{-4}$ & $6.44^{+0.38}_{-0.38}\times10^{-4}$ & $9.34\times10^{-4}$ \\
D24 & 0.2 & 26.0 & 1.0 & 3 & -13.0 & 1.00 & 1.00 & $4.69^{+0.021}_{-0.021}\times10^{-5}$ & $4.72\times10^{-5}$ & $6.23\times10^{-2}$ & $6.41^{+0.019}_{-0.019}\times10^{-2}$ & $6.47\times10^{-2}$ \\
D24 & 0.2 & 26.5 & 1.0 & 3 & -13.0 & 1.00 & 1.00 & $2.79^{+0.0052}_{-0.0052}\times10^{-4}$ & $2.70\times10^{-4}$ & $3.70\times10^{-1}$ & $3.82^{+0.0067}_{-0.0067}\times10^{-1}$ & $3.68\times10^{-1}$ \\
\tablebreak
D24 & C22+C24 & 25.5 & 1.0 & 3 & -13.0 & 1.00 & 1.00 & $2.01^{+0.14}_{-0.14}\times10^{-6}$ & $2.25\times10^{-6}$ & $2.66\times10^{-3}$ & $2.83^{+0.19}_{-0.19}\times10^{-3}$ & $3.15\times10^{-3}$ \\
D24 & A17 & 25.5 & 1.0 & 3 & -13.0 & 1.00 & 1.00 & $0^{+1.5}_{-0}\times10^{-8}$ & $<10^{-12}$ & $0$ & $0^{+2}_{-0}\times10^{-5}$ & $<10^{-9}$ \\
D24 & N20+S15 & 25.5 & 1.0 & 3 & -13.0 & 1.00 & 0.932 & $9.5^{+0.18}_{-0.18}\times10^{-6}$ & $9.66\times10^{-6}$ & $1.26\times10^{-2}$ & $1.29^{+0.016}_{-0.016}\times10^{-2}$ & $1.31\times10^{-2}$ \\
D24 & 0.2 & S24b & 1.0 & 3 & -13.0 & 1.00 & 1.00 & $1.85^{+0.015}_{-0.015}\times10^{-5}$ & $1.95\times10^{-5}$ & $2.45\times10^{-2}$ & $2.59^{+0.013}_{-0.013}\times10^{-2}$ & $2.71\times10^{-2}$ \\
D24 & 0.2 & S24a & 1.0 & 3 & -13.0 & 1.00 & 1.00 & $1.48^{+0.015}_{-0.015}\times10^{-5}$ & $1.51\times10^{-5}$ & $1.97\times10^{-2}$ & $2.02^{+0.011}_{-0.011}\times10^{-2}$ & $2.06\times10^{-2}$ \\
\enddata
\tablecomments{All incidence statistics use $R\ge2.5$~cMpc. Unsuffixed labels denote the active-fraction model, $f=f_{\rm active}$; T($t_{\rm on}$) denotes time duty, $f=f_{\rm time}$, with $t_{\rm on}$ in Myr. Simulation $n_{\ge2.5}^{\rm sim}$ uses $z=13$; $\Sigma^{\rm mid}$ multiplies that density by the comoving volume per unit solid angle and redshift for $\Delta z=1$. Simulation $\Sigma^{\rm 5s}$ uses the five-slice estimator in \S\ref{sec:headline_z13}. The $1\sigma$ simulation intervals are defined in Table~\ref{tab:sim_bubbles} and exclude cosmic variance and model uncertainty. Terminal ionized fractions use $L=50$~cMpc runs at $z=5$, or at earlier complete reionization, including the stated low-redshift UVLF extrapolations. The $C=20$ rare-tail statistics use growth intervals of at most 0.3125~Myr; halving them from 0.625~Myr leaves all five threshold counts unchanged. Analytic entries use Method 2: direct tail integration at $z=13$, continuous projection over $12.5<z<13.5$ for active fractions, and the five exact slices for time duty. The five prescription rows add matching $L=50$~cMpc runs with the same stopping rule, extending the stated efficiency prescriptions to lower redshift, and Method-2 predictions using the same luminosity-dependent photon rates. The A17 analytic tail is numerically negligible and is reported below conservative numerical thresholds, not as an observational upper limit. The fiducial model is marked with $\dagger$. Labels are defined in Table~\ref{tab:params}.}
\end{deluxetable*}

\bibliography{refs}

@article{Planck2018,
  author        = {{Planck Collaboration} and {Aghanim}, N. and {Akrami}, Y. and {Ashdown}, M. and {Aumont}, J. and {Baccigalupi}, C. and {Ballardini}, M. and {Banday}, A.~J. and {Barreiro}, R.~B. and {Bartolo}, N. and {Basak}, S. and {Battye}, R. and {Benabed}, K. and {Bernard}, J.-P. and {Bersanelli}, M. and {Bielewicz}, P. and {Bock}, J.~J. and {Bond}, J.~R. and {Borrill}, J. and {Bouchet}, F.~R. and {Boulanger}, F. and {Bucher}, M. and {Burigana}, C. and {Butler}, R.~C. and {Calabrese}, E. and {Cardoso}, J.-F. and {Carron}, J. and {Challinor}, A. and {Chiang}, H.~C. and {Chluba}, J. and {Colombo}, L.~P.~L. and {Combet}, C. and {Contreras}, D. and {Crill}, B.~P. and {Cuttaia}, F. and {de Bernardis}, P. and {de Zotti}, G. and {Delabrouille}, J. and {Delouis}, J.-M. and {Di Valentino}, E. and {Diego}, J.~M. and {Dor{\'e}}, O. and {Douspis}, M. and {Ducout}, A. and {Dupac}, X. and {Dusini}, S. and {Efstathiou}, G. and {Elsner}, F. and {En{\ss}lin}, T.~A. and {Eriksen}, H.~K. and {Fantaye}, Y. and {Farhang}, M. and {Fergusson}, J. and {Fernandez-Cobos}, R. and {Finelli}, F. and {Forastieri}, F. and {Frailis}, M. and {Fraisse}, A.~A. and {Franceschi}, E. and {Frolov}, A. and {Galeotta}, S. and {Galli}, S. and {Ganga}, K. and {G{\'e}nova-Santos}, R.~T. and {Gerbino}, M. and {Ghosh}, T. and {Gonz{\'a}lez-Nuevo}, J. and {G{\'o}rski}, K.~M. and {Gratton}, S. and {Gruppuso}, A. and {Gudmundsson}, J.~E. and {Hamann}, J. and {Handley}, W. and {Hansen}, F.~K. and {Herranz}, D. and {Hildebrandt}, S.~R. and {Hivon}, E. and {Huang}, Z. and {Jaffe}, A.~H. and {Jones}, W.~C. and {Karakci}, A. and {Keih{\"a}nen}, E. and {Keskitalo}, R. and {Kiiveri}, K. and {Kim}, J. and {Kisner}, T.~S. and {Knox}, L. and {Krachmalnicoff}, N. and {Kunz}, M. and {Kurki-Suonio}, H. and {Lagache}, G. and {Lamarre}, J.-M. and {Lasenby}, A. and {Lattanzi}, M. and {Lawrence}, C.~R. and {Le Jeune}, M. and {Lemos}, P. and {Lesgourgues}, J. and {Levrier}, F. and {Lewis}, A. and {Liguori}, M. and {Lilje}, P.~B. and {Lilley}, M. and {Lindholm}, V. and {L{\'o}pez-Caniego}, M. and {Lubin}, P.~M. and {Ma}, Y.-Z. and {Mac{\'\i}as-P{\'e}rez}, J.~F. and {Maggio}, G. and {Maino}, D. and {Mandolesi}, N. and {Mangilli}, A. and {Marcos-Caballero}, A. and {Maris}, M. and {Martin}, P.~G. and {Martinelli}, M. and {Mart{\'\i}nez-Gonz{\'a}lez}, E. and {Matarrese}, S. and {Mauri}, N. and {McEwen}, J.~D. and {Meinhold}, P.~R. and {Melchiorri}, A. and {Mennella}, A. and {Migliaccio}, M. and {Millea}, M. and {Mitra}, S. and {Miville-Desch{\^e}nes}, M.-A. and {Molinari}, D. and {Montier}, L. and {Morgante}, G. and {Moss}, A. and {Natoli}, P. and {N{\o}rgaard-Nielsen}, H.~U. and {Pagano}, L. and {Paoletti}, D. and {Partridge}, B. and {Patanchon}, G. and {Peiris}, H.~V. and {Perrotta}, F. and {Pettorino}, V. and {Piacentini}, F. and {Polastri}, L. and {Polenta}, G. and {Puget}, J.-L. and {Rachen}, J.~P. and {Reinecke}, M. and {Remazeilles}, M. and {Renzi}, A. and {Rocha}, G. and {Rosset}, C. and {Roudier}, G. and {Rubi{\~n}o-Mart{\'\i}n}, J.~A. and {Ruiz-Granados}, B. and {Salvati}, L. and {Sandri}, M. and {Savelainen}, M. and {Scott}, D. and {Shellard}, E.~P.~S. and {Sirignano}, C. and {Sirri}, G. and {Spencer}, L.~D. and {Sunyaev}, R. and {Suur-Uski}, A.-S. and {Tauber}, J.~A. and {Tavagnacco}, D. and {Tenti}, M. and {Toffolatti}, L. and {Tomasi}, M. and {Trombetti}, T. and {Valenziano}, L. and {Valiviita}, J. and {Van Tent}, B. and {Vibert}, L. and {Vielva}, P. and {Villa}, F. and {Vittorio}, N. and {Wandelt}, B.~D. and {Wehus}, I.~K. and {White}, M. and {White}, S.~D.~M. and {Zacchei}, A. and {Zonca}, A.},
  title         = {{Planck 2018 results. VI. Cosmological parameters}},
  journal       = {\aap},
  year          = 2020,
  month         = sep,
  volume        = {641},
  eid           = {A6},
  pages         = {A6},
  doi           = {10.1051/0004-6361/201833910},
  archiveprefix = {arXiv},
  eprint        = {1807.06209},
  primaryclass  = {astro-ph.CO},
  adsurl        = {https://ui.adsabs.harvard.edu/abs/2020A&A...641A...6P}
}

@article{Pagano2020,
  author        = {{Pagano}, L. and {Delouis}, J.-M. and {Mottet}, S. and {Puget}, J.-L. and {Vibert}, L.},
  title         = {{Reionization optical depth determination from Planck HFI data with ten percent accuracy}},
  journal       = {\aap},
  year          = 2020,
  month         = mar,
  volume        = {635},
  eid           = {A99},
  pages         = {A99},
  doi           = {10.1051/0004-6361/201936630},
  archiveprefix = {arXiv},
  eprint        = {1908.09856},
  primaryclass  = {astro-ph.CO},
  adsurl        = {https://ui.adsabs.harvard.edu/abs/2020A&A...635A..99P}
}

@article{Furlanetto2004,
  author        = {{Furlanetto}, Steven R. and {Zaldarriaga}, Matias and {Hernquist}, Lars},
  title         = {{The Growth of H II Regions During Reionization}},
  journal       = {\apj},
  year          = 2004,
  month         = sep,
  volume        = {613},
  number        = {1},
  pages         = {1-15},
  doi           = {10.1086/423025},
  archiveprefix = {arXiv},
  eprint        = {astro-ph/0403697},
  primaryclass  = {astro-ph},
  adsurl        = {https://ui.adsabs.harvard.edu/abs/2004ApJ...613....1F}
}

@article{McQuinn2007,
  author        = {{McQuinn}, Matthew and {Lidz}, Adam and {Zahn}, Oliver and {Dutta}, Suvendra and {Hernquist}, Lars and {Zaldarriaga}, Matias},
  title         = {{The morphology of HII regions during reionization}},
  journal       = {\mnras},
  year          = 2007,
  month         = may,
  volume        = {377},
  number        = {3},
  pages         = {1043-1063},
  doi           = {10.1111/j.1365-2966.2007.11489.x},
  archiveprefix = {arXiv},
  eprint        = {astro-ph/0610094},
  primaryclass  = {astro-ph},
  adsurl        = {https://ui.adsabs.harvard.edu/abs/2007MNRAS.377.1043M}
}

@article{Zahn2007,
  author        = {{Zahn}, Oliver and {Lidz}, Adam and {McQuinn}, Matthew and {Dutta}, Suvendra and {Hernquist}, Lars and {Zaldarriaga}, Matias and {Furlanetto}, Steven R.},
  title         = {{Simulations and Analytic Calculations of Bubble Growth during Hydrogen Reionization}},
  journal       = {\apj},
  year          = 2007,
  month         = jan,
  volume        = {654},
  number        = {1},
  pages         = {12-26},
  doi           = {10.1086/509597},
  archiveprefix = {arXiv},
  eprint        = {astro-ph/0604177},
  primaryclass  = {astro-ph},
  adsurl        = {https://ui.adsabs.harvard.edu/abs/2007ApJ...654...12Z}
}

@article{FurlanettoOh2016,
  author        = {{Furlanetto}, Steven R. and {Oh}, S. Peng},
  title         = {{Reionization through the lens of percolation theory}},
  journal       = {\mnras},
  year          = 2016,
  month         = apr,
  volume        = {457},
  number        = {2},
  pages         = {1813-1827},
  doi           = {10.1093/mnras/stw104},
  archiveprefix = {arXiv},
  eprint        = {1511.01521},
  primaryclass  = {astro-ph.CO},
  adsurl        = {https://ui.adsabs.harvard.edu/abs/2016MNRAS.457.1813F}
}

@article{Lin2016,
  author        = {{Lin}, Yin and {Oh}, S. Peng and {Furlanetto}, Steven R. and {Sutter}, P.~M.},
  title         = {{The distribution of bubble sizes during reionization}},
  journal       = {\mnras},
  year          = 2016,
  month         = sep,
  volume        = {461},
  number        = {3},
  pages         = {3361-3374},
  doi           = {10.1093/mnras/stw1542},
  archiveprefix = {arXiv},
  eprint        = {1511.01506},
  primaryclass  = {astro-ph.CO},
  adsurl        = {https://ui.adsabs.harvard.edu/abs/2016MNRAS.461.3361L}
}

@article{Giri2018,
  author        = {{Giri}, Sambit K. and {Mellema}, Garrelt and {Dixon}, Keri L. and {Iliev}, Ilian T.},
  title         = {{Bubble size statistics during reionization from 21-cm tomography}},
  journal       = {\mnras},
  year          = 2018,
  month         = jan,
  volume        = {473},
  number        = {3},
  pages         = {2949-2964},
  doi           = {10.1093/mnras/stx2539},
  archiveprefix = {arXiv},
  eprint        = {1706.00665},
  primaryclass  = {astro-ph.CO},
  adsurl        = {https://ui.adsabs.harvard.edu/abs/2018MNRAS.473.2949G}
}

@article{Finkelstein2023CEERS,
  author        = {{Finkelstein}, Steven L. and {Bagley}, Micaela B. and {Ferguson}, Henry C. and {Wilkins}, Stephen M. and {Kartaltepe}, Jeyhan S. and {Papovich}, Casey and {Yung}, L.~Y. Aaron and {Arrabal Haro}, Pablo and {Behroozi}, Peter and {Dickinson}, Mark and {Kocevski}, Dale D. and {Koekemoer}, Anton M. and {Larson}, Rebecca L. and {Le Bail}, Aur{\'e}lien and {Morales}, Alexa M. and {P{\'e}rez-Gonz{\'a}lez}, Pablo G. and {Burgarella}, Denis and {Dav{\'e}}, Romeel and {Hirschmann}, Michaela and {Somerville}, Rachel S. and {Wuyts}, Stijn and {Bromm}, Volker and {Casey}, Caitlin M. and {Fontana}, Adriano and {Fujimoto}, Seiji and {Gardner}, Jonathan P. and {Giavalisco}, Mauro and {Grazian}, Andrea and {Grogin}, Norman A. and {Hathi}, Nimish P. and {Hutchison}, Taylor A. and {Jha}, Saurabh W. and {Jogee}, Shardha and {Kewley}, Lisa J. and {Kirkpatrick}, Allison and {Long}, Arianna S. and {Lotz}, Jennifer M. and {Pentericci}, Laura and {Pierel}, Justin D.~R. and {Pirzkal}, Nor and {Ravindranath}, Swara and {Ryan}, Russell E. and {Trump}, Jonathan R. and {Yang}, Guang and {Bhatawdekar}, Rachana and {Bisigello}, Laura and {Buat}, V{\'e}ronique and {Calabr{\`o}}, Antonello and {Castellano}, Marco and {Cleri}, Nikko J. and {Cooper}, M.~C. and {Croton}, Darren and {Daddi}, Emanuele and {Dekel}, Avishai and {Elbaz}, David and {Franco}, Maximilien and {Gawiser}, Eric and {Holwerda}, Benne W. and {Huertas-Company}, Marc and {Jaskot}, Anne E. and {Leung}, Gene C.~K. and {Lucas}, Ray A. and {Mobasher}, Bahram and {Pandya}, Viraj and {Tacchella}, Sandro and {Weiner}, Benjamin J. and {Zavala}, Jorge A.},
  title         = {{CEERS Key Paper. I. An Early Look into the First 500 Myr of Galaxy Formation with JWST}},
  journal       = {\apjl},
  year          = 2023,
  month         = mar,
  volume        = {946},
  number        = {1},
  eid           = {L13},
  pages         = {L13},
  doi           = {10.3847/2041-8213/acade4},
  archiveprefix = {arXiv},
  eprint        = {2211.05792},
  primaryclass  = {astro-ph.GA},
  adsurl        = {https://ui.adsabs.harvard.edu/abs/2023ApJ...946L..13F}
}

@article{Harikane2025JWST,
  author        = {{Harikane}, Yuichi and {Inoue}, Akio K. and {Ellis}, Richard S. and {Ouchi}, Masami and {Nakazato}, Yurina and {Yoshida}, Naoki and {Ono}, Yoshiaki and {Sun}, Fengwu and {Sato}, Riku A. and {Ferrami}, Giovanni and {Fujimoto}, Seiji and {Kashikawa}, Nobunari and {McLeod}, Derek J. and {P{\'e}rez-Gonz{\'a}lez}, Pablo G. and {Sawicki}, Marcin and {Sugahara}, Yuma and {Xu}, Yi and {Yamanaka}, Satoshi and {Carnall}, Adam C. and {Cullen}, Fergus and {Dunlop}, James S. and {Egami}, Eiichi and {Grogin}, Norman and {Isobe}, Yuki and {Koekemoer}, Anton M. and {Laporte}, Nicolas and {Lee}, Chien-Hsiu and {Magee}, Dan and {Matsuo}, Hiroshi and {Matsuoka}, Yoshiki and {Mawatari}, Ken and {Nakajima}, Kimihiko and {Nakane}, Minami and {Tamura}, Yoichi and {Umeda}, Hiroya and {Yanagisawa}, Hiroto},
  title         = {{JWST, ALMA, and Keck Spectroscopic Constraints on the UV Luminosity Functions at z {\ensuremath{\sim}} 7--14: Clumpiness and Compactness of the Brightest Galaxies in the Early Universe}},
  journal       = {\apj},
  year          = 2025,
  month         = feb,
  volume        = {980},
  number        = {1},
  eid           = {138},
  pages         = {138},
  doi           = {10.3847/1538-4357/ad9b2c},
  archiveprefix = {arXiv},
  eprint        = {2406.18352},
  primaryclass  = {astro-ph.GA},
  adsurl        = {https://ui.adsabs.harvard.edu/abs/2025ApJ...980..138H}
}

@article{Bouwens2021,
  author        = {{Bouwens}, R.~J. and {Oesch}, P.~A. and {Stefanon}, M. and {Illingworth}, G. and {Labb{\'e}}, I. and {Reddy}, N. and {Atek}, H. and {Montes}, M. and {Naidu}, R. and {Nanayakkara}, T. and {Nelson}, E. and {Wilkins}, S.},
  title         = {{New Determinations of the UV Luminosity Functions from $z\sim9$ to 2 Show a Remarkable Consistency with Halo Growth and a Constant Star Formation Efficiency}},
  journal       = {\aj},
  year          = 2021,
  month         = aug,
  volume        = {162},
  number        = {2},
  eid           = {47},
  pages         = {47},
  doi           = {10.3847/1538-3881/abf83e},
  archiveprefix = {arXiv},
  eprint        = {2102.07775},
  primaryclass  = {astro-ph.GA},
  adsurl        = {https://ui.adsabs.harvard.edu/abs/2021AJ....162...47B}
}

@article{Donnan2024PRIMER,
  author        = {{Donnan}, C.~T. and {McLure}, R.~J. and {Dunlop}, J.~S. and {McLeod}, D.~J. and {Magee}, D. and {Arellano-C{\'o}rdova}, K.~Z. and {Barrufet}, L. and {Begley}, R. and {Bowler}, R.~A.~A. and {Carnall}, A.~C. and {Cullen}, F. and {Ellis}, R.~S. and {Fontana}, A. and {Illingworth}, G.~D. and {Grogin}, N.~A. and {Hamadouche}, M.~L. and {Koekemoer}, A.~M. and {Liu}, F.-Y. and {Mason}, C. and {Santini}, P. and {Stanton}, T.~M.},
  title         = {{JWST PRIMER: a new multifield determination of the evolving galaxy UV luminosity function at redshifts z ≃ 9 - 15}},
  journal       = {\mnras},
  year          = 2024,
  month         = sep,
  volume        = {533},
  number        = {3},
  pages         = {3222-3237},
  doi           = {10.1093/mnras/stae2037},
  archiveprefix = {arXiv},
  eprint        = {2403.03171},
  primaryclass  = {astro-ph.GA},
  adsurl        = {https://ui.adsabs.harvard.edu/abs/2024MNRAS.533.3222D}
}

@article{Mason2018,
  author        = {{Mason}, Charlotte A. and {Treu}, Tommaso and {Dijkstra}, Mark and {Mesinger}, Andrei and {Trenti}, Michele and {Pentericci}, Laura and {de Barros}, Stephane and {Vanzella}, Eros},
  title         = {{The Universe Is Reionizing at z {\ensuremath{\sim}} 7: Bayesian Inference of the IGM Neutral Fraction Using Ly{\ensuremath{\alpha}} Emission from Galaxies}},
  journal       = {\apj},
  year          = 2018,
  month         = mar,
  volume        = {856},
  number        = {1},
  eid           = {2},
  pages         = {2},
  doi           = {10.3847/1538-4357/aab0a7},
  archiveprefix = {arXiv},
  eprint        = {1709.05356},
  primaryclass  = {astro-ph.CO},
  adsurl        = {https://ui.adsabs.harvard.edu/abs/2018ApJ...856....2M}
}

@article{Tang2024,
  author        = {{Tang}, Mengtao and {Stark}, Daniel P. and {Topping}, Michael W. and {Mason}, Charlotte and {Ellis}, Richard S.},
  title         = {{JWST/NIRSpec Observations of Lyman $\alpha$ Emission in Star-forming Galaxies at $6.5 \lesssim z \lesssim 13$}},
  journal       = {\apj},
  year          = 2024,
  month         = nov,
  volume        = {975},
  number        = {2},
  eid           = {208},
  pages         = {208},
  doi           = {10.3847/1538-4357/ad7eb7},
  archiveprefix = {arXiv},
  eprint        = {2408.01507},
  primaryclass  = {astro-ph.GA},
  adsurl        = {https://ui.adsabs.harvard.edu/abs/2024ApJ...975..208T}
}

@article{Endsley2023JADES,
  author        = {{Endsley}, Ryan and {Stark}, Daniel P. and {Whitler}, Lily and {Topping}, Michael W. and {Johnson}, Benjamin D. and {Robertson}, Brant and {Tacchella}, Sandro and {Alberts}, Stacey and {Baker}, William M. and {Bhatawdekar}, Rachana and {Boyett}, Kristan and {Bunker}, Andrew J. and {Cameron}, Alex J. and {Carniani}, Stefano and {Charlot}, Stephane and {Chen}, Zuyi and {Chevallard}, Jacopo and {Curtis-Lake}, Emma and {Danhaive}, A. Lola and {Egami}, Eiichi and {Eisenstein}, Daniel J. and {Hainline}, Kevin and {Helton}, Jakob M. and {Ji}, Zhiyuan and {Looser}, Tobias J. and {Maiolino}, Roberto and {Nelson}, Erica and {Pusk{\'a}s}, D{\'a}vid and {Rieke}, George and {Rieke}, Marcia and {Rix}, Hans-Walter and {Sandles}, Lester and {Saxena}, Aayush and {Simmonds}, Charlotte and {Smit}, Renske and {Sun}, Fengwu and {Williams}, Christina C. and {Willmer}, Christopher N.~A. and {Willott}, Chris and {Witstok}, Joris},
  title         = {{The star-forming and ionizing properties of dwarf $z\sim6$--9 galaxies in JADES: insights on bursty star formation and ionized bubble growth}},
  journal       = {\mnras},
  year          = 2024,
  month         = sep,
  volume        = {533},
  number        = {1},
  pages         = {1111-1142},
  doi           = {10.1093/mnras/stae1857},
  archiveprefix = {arXiv},
  eprint        = {2306.05295},
  primaryclass  = {astro-ph.GA},
  adsurl        = {https://ui.adsabs.harvard.edu/abs/2024MNRAS.533.1111E}
}

@article{Endsley2024Burstiness,
  author        = {{Endsley}, Ryan and {Chisholm}, John and {Stark}, Daniel P. and {Topping}, Michael W. and {Whitler}, Lily},
  title         = {{The Burstiness of Star Formation at z {\ensuremath{\sim}} 6: A Huge Diversity in the Recent Star Formation Histories of Very UV-faint Galaxies}},
  journal       = {\apj},
  year          = 2025,
  month         = jul,
  volume        = {987},
  number        = {2},
  eid           = {189},
  pages         = {189},
  doi           = {10.3847/1538-4357/addc74},
  archiveprefix = {arXiv},
  eprint        = {2410.01905},
  primaryclass  = {astro-ph.GA},
  adsurl        = {https://ui.adsabs.harvard.edu/abs/2025ApJ...987..189E}
}

@article{Stiavelli2026,
  author        = {{Stiavelli}, Massimo and {Ricotti}, Massimo},
  title         = {{How Bursty is Star Formation at z > 5?}},
  journal       = {\apj},
  year          = 2026,
  month         = mar,
  volume        = {1000},
  number        = {1},
  eid           = {96},
  pages         = {96},
  doi           = {10.3847/1538-4357/ae486c},
  archiveprefix = {arXiv},
  eprint        = {2602.16706},
  primaryclass  = {astro-ph.GA},
  adsurl        = {https://ui.adsabs.harvard.edu/abs/2026ApJ..1000...96S}
}

@article{Munoz2026Burstiness,
  author        = {{Mu{\~n}oz}, Julian B. and {Chisholm}, John and {Sun}, Guochao and {Samuel}, Jenna and {Mirocha}, Jordan and {Bregou}, Emily and {Venditti}, Alessandra and {Qezlou}, Mahdi and {Simmonds}, Charlotte and {Endsley}, Ryan},
  title         = {{Relatively fast and reasonably furious: evidence for increased burstiness in smaller haloes at cosmic dawn}},
  journal       = {\mnras},
  year          = 2026,
  month         = apr,
  volume        = {547},
  number        = {4},
  eid           = {stag415},
  pages         = {stag415},
  doi           = {10.1093/mnras/stag415},
  archiveprefix = {arXiv},
  eprint        = {2601.07912},
  primaryclass  = {astro-ph.GA},
  adsurl        = {https://ui.adsabs.harvard.edu/abs/2026MNRAS.547ag415M}
}

@article{Dalmasso2024Clustering,
  author        = {{Dalmasso}, Nicol{\`o} and {Leethochawalit}, Nicha and {Trenti}, Michele and {Boyett}, Kristan},
  title         = {{Galaxy clustering at cosmic dawn from JWST/NIRCam observations to redshift z {\ensuremath{\sim}} 11}},
  journal       = {\mnras},
  year          = 2024,
  month         = sep,
  volume        = {533},
  number        = {2},
  pages         = {2391-2398},
  doi           = {10.1093/mnras/stae2006},
  archiveprefix = {arXiv},
  eprint        = {2402.18052},
  primaryclass  = {astro-ph.GA},
  adsurl        = {https://ui.adsabs.harvard.edu/abs/2024MNRAS.533.2391D}
}

@article{Witstok2025JADESLAEs,
  author        = {{Witstok}, Joris and {Maiolino}, Roberto and {Smit}, Renske and {Jones}, Gareth C. and {Bunker}, Andrew J. and {Helton}, Jakob M. and {Johnson}, Benjamin D. and {Tacchella}, Sandro and {Saxena}, Aayush and {Arribas}, Santiago and {Bhatawdekar}, Rachana and {Boyett}, Kristan and {Cameron}, Alex J. and {Cargile}, Phillip A. and {Carniani}, Stefano and {Charlot}, St{\'e}phane and {Chevallard}, Jacopo and {Curti}, Mirko and {Curtis-Lake}, Emma and {D'Eugenio}, Francesco and {Eisenstein}, Daniel J. and {Hainline}, Kevin and {Hausen}, Ryan and {Kumari}, Nimisha and {Laseter}, Isaac and {Maseda}, Michael V. and {Rieke}, Marcia and {Robertson}, Brant and {Scholtz}, Jan and {Shivaei}, Irene and {Williams}, Christina C. and {Willmer}, Christopher N.~A. and {Willott}, Chris},
  title         = {{JADES: primaeval Lyman-{\ensuremath{\alpha}} emitting galaxies reveal early sites of reionization out to redshift z {\ensuremath{\sim}} 9}},
  journal       = {\mnras},
  year          = 2025,
  month         = jan,
  volume        = {536},
  number        = {1},
  pages         = {27-50},
  doi           = {10.1093/mnras/stae2535},
  archiveprefix = {arXiv},
  eprint        = {2404.05724},
  primaryclass  = {astro-ph.GA},
  adsurl        = {https://ui.adsabs.harvard.edu/abs/2025MNRAS.536...27W}
}

@article{Witstok2025Nature,
  author        = {{Witstok}, Joris and {Jakobsen}, Peter and {Maiolino}, Roberto and {Helton}, Jakob M. and {Johnson}, Benjamin D. and {Robertson}, Brant E. and {Tacchella}, Sandro and {Cameron}, Alex J. and {Smit}, Renske and {Bunker}, Andrew J. and {Saxena}, Aayush and {Sun}, Fengwu and {Alberts}, Stacey and {Arribas}, Santiago and {Baker}, William M. and {Bhatawdekar}, Rachana and {Boyett}, Kristan and {Cargile}, Phillip A. and {Carniani}, Stefano and {Charlot}, St{\'e}phane and {Chevallard}, Jacopo and {Curti}, Mirko and {Curtis-Lake}, Emma and {D'Eugenio}, Francesco and {Eisenstein}, Daniel J. and {Hainline}, Kevin N. and {Jones}, Gareth C. and {Kumari}, Nimisha and {Maseda}, Michael V. and {P{\'e}rez-Gonz{\'a}lez}, Pablo G. and {Rinaldi}, Pierluigi and {Scholtz}, Jan and {{\"U}bler}, Hannah and {Williams}, Christina C. and {Willmer}, Christopher N.~A. and {Willott}, Chris and {Zhu}, Yongda},
  title         = {{Witnessing the onset of reionization through Lyman-{\ensuremath{\alpha}} emission at redshift 13}},
  journal       = {\nat},
  year          = 2025,
  month         = mar,
  volume        = {639},
  number        = {8056},
  pages         = {897-901},
  doi           = {10.1038/s41586-025-08779-5},
  archiveprefix = {arXiv},
  eprint        = {2408.16608},
  primaryclass  = {astro-ph.GA},
  adsurl        = {https://ui.adsabs.harvard.edu/abs/2025Natur.639..897W}
}

@article{Pawlik2009,
  author        = {{Pawlik}, Andreas H. and {Schaye}, Joop and {van Scherpenzeel}, Eveline},
  title         = {{Keeping the Universe ionized: photoheating and the clumping factor of the high-redshift intergalactic medium}},
  journal       = {\mnras},
  year          = 2009,
  month         = apr,
  volume        = {394},
  number        = {4},
  pages         = {1812-1824},
  doi           = {10.1111/j.1365-2966.2009.14486.x},
  archiveprefix = {arXiv},
  eprint        = {0807.3963},
  primaryclass  = {astro-ph},
  adsurl        = {https://ui.adsabs.harvard.edu/abs/2009MNRAS.394.1812P}
}

@article{Finlator2012,
  author        = {{Finlator}, Kristian and {Oh}, S. Peng and {{\"O}zel}, Feryal and {Dav{\'e}}, Romeel},
  title         = {{Gas clumping in self-consistent reionization models}},
  journal       = {\mnras},
  year          = 2012,
  month         = dec,
  volume        = {427},
  number        = {3},
  pages         = {2464-2479},
  doi           = {10.1111/j.1365-2966.2012.22114.x},
  archiveprefix = {arXiv},
  eprint        = {1209.2489},
  primaryclass  = {astro-ph.CO},
  adsurl        = {https://ui.adsabs.harvard.edu/abs/2012MNRAS.427.2464F}
}

@article{Sobacchi2014,
  author        = {{Sobacchi}, Emanuele and {Mesinger}, Andrei},
  title         = {{Inhomogeneous recombinations during cosmic reionization}},
  journal       = {\mnras},
  year          = 2014,
  month         = may,
  volume        = {440},
  number        = {2},
  pages         = {1662-1673},
  doi           = {10.1093/mnras/stu377},
  archiveprefix = {arXiv},
  eprint        = {1402.2298},
  primaryclass  = {astro-ph.CO},
  adsurl        = {https://ui.adsabs.harvard.edu/abs/2014MNRAS.440.1662S}
}

@article{Totani2006,
  author        = {{Totani}, Tomonori and {Kawai}, Nobuyuki and {Kosugi}, George and {Aoki}, Kentaro and {Yamada}, Toru and {Iye}, Masanori and {Ohta}, Kouji and {Hattori}, Takashi},
  title         = {{Implications for Cosmic Reionization from the Optical Afterglow Spectrum of the Gamma-Ray Burst 050904 at z = 6.3$^{*}$}},
  journal       = {\pasj},
  year          = 2006,
  month         = jun,
  volume        = {58},
  number        = {3},
  pages         = {485-498},
  doi           = {10.1093/pasj/58.3.485},
  archiveprefix = {arXiv},
  eprint        = {astro-ph/0512154},
  primaryclass  = {astro-ph},
  adsurl        = {https://ui.adsabs.harvard.edu/abs/2006PASJ...58..485T}
}

@article{Davies2018,
  author        = {{Davies}, Frederick B. and {Hennawi}, Joseph F. and {Ba{\~n}ados}, Eduardo and {Luki{\'c}}, Zarija and {Decarli}, Roberto and {Fan}, Xiaohui and {Farina}, Emanuele P. and {Mazzucchelli}, Chiara and {Rix}, Hans-Walter and {Venemans}, Bram P. and {Walter}, Fabian and {Wang}, Feige and {Yang}, Jinyi},
  title         = {{Quantitative Constraints on the Reionization History from the IGM Damping Wing Signature in Two Quasars at z > 7}},
  journal       = {\apj},
  year          = 2018,
  month         = sep,
  volume        = {864},
  number        = {2},
  eid           = {142},
  pages         = {142},
  doi           = {10.3847/1538-4357/aad6dc},
  archiveprefix = {arXiv},
  eprint        = {1802.06066},
  primaryclass  = {astro-ph.CO},
  adsurl        = {https://ui.adsabs.harvard.edu/abs/2018ApJ...864..142D}
}

@article{Wang2020,
  author        = {{Wang}, Feige and {Davies}, Frederick B. and {Yang}, Jinyi and {Hennawi}, Joseph F. and {Fan}, Xiaohui and {Barth}, Aaron J. and {Jiang}, Linhua and {Wu}, Xue-Bing and {Mudd}, Dale M. and {Ba{\~n}ados}, Eduardo and {Bian}, Fuyan and {Decarli}, Roberto and {Eilers}, Anna-Christina and {Farina}, Emanuele Paolo and {Venemans}, Bram and {Walter}, Fabian and {Yue}, Minghao},
  title         = {{A Significantly Neutral Intergalactic Medium Around the Luminous z=7 Quasar J0252-0503}},
  journal       = {\apj},
  year          = 2020,
  volume        = {896},
  pages         = {23},
  doi           = {10.3847/1538-4357/ab8c45},
  archiveprefix = {arXiv},
  eprint        = {2004.10877},
  primaryclass  = {astro-ph.GA}
}

@article{Greig2022,
  author        = {{Greig}, Bradley and {Mesinger}, Andrei and {Davies}, Frederick B. and {Wang}, Feige and {Yang}, Jinyi and {Hennawi}, Joseph F.},
  title         = {{IGM damping wing constraints on reionization from covariance reconstruction of two z {\ensuremath{\gtrsim}} 7 QSOs}},
  journal       = {\mnras},
  year          = 2022,
  month         = jun,
  volume        = {512},
  number        = {4},
  pages         = {5390-5403},
  doi           = {10.1093/mnras/stac825},
  archiveprefix = {arXiv},
  eprint        = {2112.04091},
  primaryclass  = {astro-ph.CO},
  adsurl        = {https://ui.adsabs.harvard.edu/abs/2022MNRAS.512.5390G}
}

@article{Durovcikova2024,
  author        = {{{\v{D}}urov{\v{c}}{\'\i}kov{\'a}}, Dominika and {Eilers}, Anna-Christina and {Chen}, Huanqing and {Satyavolu}, Sindhu and {Kulkarni}, Girish and {Simcoe}, Robert A. and {Keating}, Laura C. and {Haehnelt}, Martin G. and {Ba{\~n}ados}, Eduardo},
  title         = {{Chronicling the Reionization History at $6\lesssim z\lesssim7$ with Emergent Quasar Damping Wings}},
  journal       = {\apj},
  year          = 2024,
  volume        = {969},
  pages         = {162},
  doi           = {10.3847/1538-4357/ad4888},
  archiveprefix = {arXiv},
  eprint        = {2401.10328},
  primaryclass  = {astro-ph.CO}
}

@article{Umeda2025,
  author        = {{Umeda}, Hiroya and {Ouchi}, Masami and {Kageura}, Yuta and {Harikane}, Yuichi and {Nakane}, Minami and {Thai}, Tran Thi and {Nakajima}, Kimihiko},
  title         = {{Probing the Cosmic Reionization History with JWST: Gunn--Peterson and Ly{\ensuremath{\alpha}} Damping Wing Absorption at 4.5 < z < 13}},
  journal       = {\apj},
  year          = 2026,
  month         = jan,
  volume        = {997},
  number        = {1},
  eid           = {86},
  pages         = {86},
  doi           = {10.3847/1538-4357/ae232b},
  archiveprefix = {arXiv},
  eprint        = {2504.04683},
  primaryclass  = {astro-ph.GA},
  adsurl        = {https://ui.adsabs.harvard.edu/abs/2026ApJ...997...86U}
}

@article{Hoag2019,
  author        = {{Hoag}, A. and {Brada{\v{c}}}, M. and {Huang}, K. and {Mason}, C. and {Treu}, T. and {Schmidt}, K.~B. and {Trenti}, M. and {Strait}, V. and {Lemaux}, B.~C. and {Finney}, E.~Q. and {Paddock}, M.},
  title         = {{Constraining the Neutral Fraction of Hydrogen in the IGM at Redshift 7.5}},
  journal       = {\apj},
  year          = 2019,
  month         = jun,
  volume        = {878},
  number        = {1},
  eid           = {12},
  pages         = {12},
  doi           = {10.3847/1538-4357/ab1de7},
  archiveprefix = {arXiv},
  eprint        = {1901.09001},
  primaryclass  = {astro-ph.GA},
  adsurl        = {https://ui.adsabs.harvard.edu/abs/2019ApJ...878...12H}
}

@article{Mason2019,
  author        = {{Mason}, Charlotte A. and {Fontana}, Adriano and {Treu}, Tommaso and {Schmidt}, Kasper B. and {Hoag}, Austin and {Abramson}, Louis and {Amorin}, Ricardo and {Brada{\v{c}}}, Maru{\v{s}}a and {Guaita}, Lucia and {Jones}, Tucker and {Henry}, Alaina and {Malkan}, Matthew A. and {Pentericci}, Laura and {Trenti}, Michele and {Vanzella}, Eros},
  title         = {{Inferences on the timeline of reionization at z {\ensuremath{\sim}} 8 from the KMOS Lens-Amplified Spectroscopic Survey}},
  journal       = {\mnras},
  year          = 2019,
  month         = may,
  volume        = {485},
  number        = {3},
  pages         = {3947-3969},
  doi           = {10.1093/mnras/stz632},
  archiveprefix = {arXiv},
  eprint        = {1901.11045},
  primaryclass  = {astro-ph.CO},
  adsurl        = {https://ui.adsabs.harvard.edu/abs/2019MNRAS.485.3947M}
}

@article{Bolan2022,
  author        = {{Bolan}, Patricia and {Lemaux}, Brian C. and {Mason}, Charlotte and {Brada{\v{c}}}, Maru{\v{s}}a and {Treu}, Tommaso and {Strait}, Victoria and {Pelliccia}, Debora and {Pentericci}, Laura and {Malkan}, Matthew},
  title         = {{Inferring the intergalactic medium neutral fraction at $z\sim6$--8 with low-luminosity Lyman break galaxies}},
  journal       = {\mnras},
  year          = 2022,
  month         = dec,
  volume        = {517},
  number        = {3},
  pages         = {3263-3274},
  doi           = {10.1093/mnras/stac1963},
  archiveprefix = {arXiv},
  eprint        = {2111.14912},
  primaryclass  = {astro-ph.GA},
  adsurl        = {https://ui.adsabs.harvard.edu/abs/2022MNRAS.517.3263B}
}

@article{McGreer2015,
  author        = {{McGreer}, Ian D. and {Mesinger}, Andrei and {D'Odorico}, Valentina},
  title         = {{Model-independent evidence in favour of an end to reionization by z {\ensuremath{\approx}} 6}},
  journal       = {\mnras},
  year          = 2015,
  month         = feb,
  volume        = {447},
  number        = {1},
  pages         = {499-505},
  doi           = {10.1093/mnras/stu2449},
  archiveprefix = {arXiv},
  eprint        = {1411.5375},
  primaryclass  = {astro-ph.CO},
  adsurl        = {https://ui.adsabs.harvard.edu/abs/2015MNRAS.447..499M}
}

@article{Davies2026,
  author        = {{Davies}, Frederick B. and {Bosman}, Sarah E.~I. and {D'Odorico}, Valentina and {Campo}, Sofia and {Mesinger}, Andrei and {Qin}, Yuxiang and {Becker}, George D. and {Ba{\~n}ados}, Eduardo and {Chen}, Huanqing and {Cristiani}, Stefano and {Fan}, Xiaohui and {Gallerani}, Simona and {Haehnelt}, Martin G. and {Keating}, Laura C. and {Lai}, Samuel and {Ryan-Weber}, Emma and {Wang}, Feige and {Yang}, Jinyi and {Zhu}, Yongda},
  title         = {{Updated dark pixel fraction constraints on reionization's end from the Lyman-series forests of XQR{\ensuremath{-}}30}},
  journal       = {\mnras},
  year          = 2026,
  month         = jan,
  volume        = {545},
  number        = {2},
  eid           = {staf1862},
  pages         = {staf1862},
  doi           = {10.1093/mnras/staf1862},
  archiveprefix = {arXiv},
  eprint        = {2510.25829},
  primaryclass  = {astro-ph.CO},
  adsurl        = {https://ui.adsabs.harvard.edu/abs/2026MNRAS.545f1862D}
}

@article{Atek2024,
  author        = {{Atek}, Hakim and {Labb{\'e}}, Ivo and {Furtak}, Lukas J. and {Chemerynska}, Iryna and {Fujimoto}, Seiji and {Setton}, David J. and {Miller}, Tim B. and {Oesch}, Pascal and {Bezanson}, Rachel and {Price}, Sedona H. and {Dayal}, Pratika and {Zitrin}, Adi and {Kokorev}, Vasily and {Weaver}, John R. and {Brammer}, Gabriel and {Dokkum}, Pieter van and {Williams}, Christina C. and {Cutler}, Sam E. and {Feldmann}, Robert and {Fudamoto}, Yoshinobu and {Greene}, Jenny E. and {Leja}, Joel and {Maseda}, Michael V. and {Muzzin}, Adam and {Pan}, Richard and {Papovich}, Casey and {Nelson}, Erica J. and {Nanayakkara}, Themiya and {Stark}, Daniel P. and {Stefanon}, Mauro and {Suess}, Katherine A. and {Wang}, Bingjie and {Whitaker}, Katherine E.},
  title         = {{Most of the photons that reionized the Universe came from dwarf galaxies}},
  journal       = {\nat},
  year          = 2024,
  month         = feb,
  volume        = {626},
  number        = {8001},
  pages         = {975-978},
  doi           = {10.1038/s41586-024-07043-6},
  archiveprefix = {arXiv},
  eprint        = {2308.08540},
  primaryclass  = {astro-ph.GA},
  adsurl        = {https://ui.adsabs.harvard.edu/abs/2024Natur.626..975A}
}

@article{Maiolino2024GNz11,
  author        = {{Maiolino}, Roberto and {Scholtz}, Jan and {Witstok}, Joris and {Carniani}, Stefano and {D'Eugenio}, Francesco and {de Graaff}, Anna and {{\"U}bler}, Hannah and {Tacchella}, Sandro and {Curtis-Lake}, Emma and {Arribas}, Santiago and {Bunker}, Andrew and {Charlot}, St{\'e}phane and {Chevallard}, Jacopo and {Curti}, Mirko and {Looser}, Tobias J. and {Maseda}, Michael V. and {Rawle}, Timothy D. and {Rodr{\'\i}guez del Pino}, Bruno and {Willott}, Chris J. and {Egami}, Eiichi and {Eisenstein}, Daniel J. and {Hainline}, Kevin N. and {Robertson}, Brant and {Williams}, Christina C. and {Willmer}, Christopher N.~A. and {Baker}, William M. and {Boyett}, Kristan and {DeCoursey}, Christa and {Fabian}, Andrew C. and {Helton}, Jakob M. and {Ji}, Zhiyuan and {Jones}, Gareth C. and {Kumari}, Nimisha and {Laporte}, Nicolas and {Nelson}, Erica J. and {Perna}, Michele and {Sandles}, Lester and {Shivaei}, Irene and {Sun}, Fengwu},
  title         = {{A small and vigorous black hole in the early Universe}},
  journal       = {\nat},
  year          = 2024,
  month         = mar,
  volume        = {627},
  number        = {8002},
  pages         = {59-63},
  doi           = {10.1038/s41586-024-07052-5},
  archiveprefix = {arXiv},
  eprint        = {2305.12492},
  primaryclass  = {astro-ph.GA},
  adsurl        = {https://ui.adsabs.harvard.edu/abs/2024Natur.627...59M}
}

@article{Looser2024,
  author        = {{Looser}, Tobias J. and {D'Eugenio}, Francesco and {Maiolino}, Roberto and {Witstok}, Joris and {Sandles}, Lester and {Curtis-Lake}, Emma and {Chevallard}, Jacopo and {Tacchella}, Sandro and {Johnson}, Benjamin D. and {Baker}, William M. and {Suess}, Katherine A. and {Carniani}, Stefano and {Ferruit}, Pierre and {Arribas}, Santiago and {Bonaventura}, Nina and {Bunker}, Andrew J. and {Cameron}, Alex J. and {Charlot}, Stephane and {Curti}, Mirko and {de Graaff}, Anna and {Maseda}, Michael V. and {Rawle}, Tim and {Rix}, Hans-Walter and {Del Pino}, Bruno Rodr{\'\i}guez and {Smit}, Renske and {{\"U}bler}, Hannah and {Willott}, Chris and {Alberts}, Stacey and {Egami}, Eiichi and {Eisenstein}, Daniel J. and {Endsley}, Ryan and {Hausen}, Ryan and {Rieke}, Marcia and {Robertson}, Brant and {Shivaei}, Irene and {Williams}, Christina C. and {Boyett}, Kristan and {Chen}, Zuyi and {Ji}, Zhiyuan and {Jones}, Gareth C. and {Kumari}, Nimisha and {Nelson}, Erica and {Perna}, Michele and {Saxena}, Aayush and {Scholtz}, Jan},
  title         = {{A recently quenched galaxy 700 million years after the Big Bang}},
  journal       = {\nat},
  year          = 2024,
  month         = may,
  volume        = {629},
  number        = {8010},
  pages         = {53-57},
  doi           = {10.1038/s41586-024-07227-0},
  archiveprefix = {arXiv},
  eprint        = {2302.14155},
  primaryclass  = {astro-ph.GA},
  adsurl        = {https://ui.adsabs.harvard.edu/abs/2024Natur.629...53L}
}

@article{Kannan2022THESAN,
  author        = {{Kannan}, R. and {Garaldi}, E. and {Smith}, A. and {Pakmor}, R. and {Springel}, V. and {Vogelsberger}, M. and {Hernquist}, L.},
  title         = {{Introducing the THESAN project: radiation-magnetohydrodynamic simulations of the epoch of reionization}},
  journal       = {\mnras},
  year          = 2022,
  month         = apr,
  volume        = {511},
  number        = {3},
  pages         = {4005-4030},
  doi           = {10.1093/mnras/stab3710},
  archiveprefix = {arXiv},
  eprint        = {2110.00584},
  primaryclass  = {astro-ph.GA},
  adsurl        = {https://ui.adsabs.harvard.edu/abs/2022MNRAS.511.4005K}
}

@article{Ocvirk2020,
  author        = {{Ocvirk}, Pierre and {Aubert}, Dominique and {Sorce}, Jenny G. and {Shapiro}, Paul R. and {Deparis}, Nicolas and {Dawoodbhoy}, Taha and {Lewis}, Joseph and {Teyssier}, Romain and {Yepes}, Gustavo and {Gottl{\"o}ber}, Stefan and {Ahn}, Kyungjin and {Iliev}, Ilian T. and {Hoffman}, Yehuda},
  title         = {{Cosmic Dawn II (CoDa II): a new radiation-hydrodynamics simulation of the self-consistent coupling of galaxy formation and reionization}},
  journal       = {\mnras},
  year          = 2020,
  month         = aug,
  volume        = {496},
  number        = {4},
  pages         = {4087-4107},
  doi           = {10.1093/mnras/staa1266},
  archiveprefix = {arXiv},
  eprint        = {1811.11192},
  primaryclass  = {astro-ph.GA},
  adsurl        = {https://ui.adsabs.harvard.edu/abs/2020MNRAS.496.4087O}
}

@article{Garaldi2024,
  author        = {{Garaldi}, Enrico and {Kannan}, Rahul and {Smith}, Aaron and {Borrow}, Josh and {Vogelsberger}, Mark and {Pakmor}, R{\"u}diger and {Springel}, Volker and {Hernquist}, Lars and {Gal{\'a}rraga-Espinosa}, Daniela and {Yeh}, Jessica Y.-C. and {Shen}, Xuejian and {Xu}, Clara and {Neyer}, Meredith and {Spina}, Benedetta and {Almualla}, Mouza and {Zhao}, Yu},
  title         = {{The THESAN project: public data release of radiation-hydrodynamic simulations matching reionization-era JWST observations}},
  journal       = {\mnras},
  year          = 2024,
  month         = jun,
  volume        = {530},
  number        = {4},
  pages         = {3765-3786},
  doi           = {10.1093/mnras/stae839},
  archiveprefix = {arXiv},
  eprint        = {2309.06475},
  primaryclass  = {astro-ph.CO},
  adsurl        = {https://ui.adsabs.harvard.edu/abs/2024MNRAS.530.3765G}
}

@article{Chisholm2022,
  author = {{Chisholm}, J. and {Saldana-Lopez}, A. and {Flury}, S. and others},
  title = {{The far-ultraviolet continuum slope as a Lyman Continuum escape estimator at high redshift}},
  journal = {\mnras}, year = {2022}, volume = {517}, pages = {5104--5120},
  doi = {10.1093/mnras/stac2874}, eprint = {2207.05771}, archiveprefix = {arXiv}
}

@article{Cullen2024,
  author = {{Cullen}, F. and {McLeod}, D.~J. and {McLure}, R.~J. and others},
  title = {{The ultraviolet continuum slopes of high-redshift galaxies: evidence for the emergence of dust-free stellar populations at z > 10}},
  journal = {\mnras}, year = {2024}, volume = {531}, pages = {997--1020},
  doi = {10.1093/mnras/stae1211}
}

@article{Anderson2017,
  author = {{Anderson}, Lauren and {Governato}, F. and {Karcher}, M. and {Quinn}, T. and {Wadsley}, J.},
  title = {{The little Galaxies that could (reionize the universe): predicting faint end slopes \& escape fractions at z > 4}},
  journal = {\mnras}, year = {2017}, volume = {468}, pages = {4077--4092},
  doi = {10.1093/mnras/stx709}
}

@article{Shibuya2015,
  author = {{Shibuya}, Takatoshi and {Ouchi}, Masami and {Harikane}, Yuichi},
  title = {{Morphologies of $\sim$190,000 Galaxies at z = 0--10 Revealed with HST Legacy Data. I. Size Evolution}},
  journal = {\apjs}, year = {2015}, volume = {219}, pages = {15},
  eprint = {1503.07481}, archiveprefix = {arXiv}, url = {https://arxiv.org/abs/1503.07481},
  doi = {10.1088/0067-0049/219/2/15}
}

@article{Shuntov2025,
  author = {{Shuntov}, M. and {Ilbert}, O. and {Toft}, S. and others},
  title = {{COSMOS-Web: stellar mass assembly in relation to dark matter halos across $0.2<z<12$ of cosmic history}},
  journal = {\aap}, year = {2025}, volume = {695}, pages = {A20},
  doi = {10.1051/0004-6361/202452570}, eprint = {2410.08290}, archiveprefix = {arXiv}
}

@article{Simmonds2024MassComplete,
  author = {{Simmonds}, C. and {Tacchella}, S. and {Hainline}, K. and {Johnson}, B.~D. and {Pusk{\'a}s}, D. and {Robertson}, B. and others},
  title = {{Ionizing properties of galaxies in JADES for a stellar mass complete sample: resolving the cosmic ionizing photon budget crisis at the Epoch of Reionization}},
  journal = {\mnras}, year = {2024}, volume = {535}, pages = {2998--3019},
  doi = {10.1093/mnras/stae2537}, eprint = {2409.01286}, archiveprefix = {arXiv}
}

@article{Simmonds2024,
  author        = {{Simmonds}, C. and {Tacchella}, S. and {Hainline}, K. and {Johnson}, B.~D. and {McClymont}, W. and {Robertson}, B. and {Saxena}, A. and {Sun}, F. and {Witten}, C. and {Baker}, W.~M. and {Bhatawdekar}, R. and {Boyett}, K. and {Bunker}, A.~J. and {Charlot}, S. and {Curtis-Lake}, E. and {Egami}, E. and {Eisenstein}, D.~J. and {Hausen}, R. and {Maiolino}, R. and {Maseda}, M.~V. and {Scholtz}, J. and {Williams}, C.~C. and {Willott}, C. and {Witstok}, J.},
  title         = {{Low-mass bursty galaxies in JADES efficiently produce ionizing photons and could represent the main drivers of reionization}},
  journal       = {\mnras},
  year          = 2024,
  month         = jan,
  volume        = {527},
  number        = {3},
  pages         = {6139-6157},
  doi           = {10.1093/mnras/stad3605},
  archiveprefix = {arXiv},
  eprint        = {2310.01112},
  primaryclass  = {astro-ph.GA},
  adsurl        = {https://ui.adsabs.harvard.edu/abs/2024MNRAS.527.6139S}
}

@article{Mascia2023,
  author        = {{Mascia}, S. and {Pentericci}, L. and {Calabr{\`o}}, A. and {Treu}, T. and {Santini}, P. and {Yang}, L. and {Napolitano}, L. and {Roberts-Borsani}, G. and {Bergamini}, P. and {Grillo}, C. and {Rosati}, P. and {Vulcani}, B. and {Castellano}, M. and {Boyett}, K. and {Fontana}, A. and {Glazebrook}, K. and {Henry}, A. and {Mason}, C. and {Merlin}, E. and {Morishita}, T. and {Nanayakkara}, T. and {Paris}, D. and {Roy}, N. and {Williams}, H. and {Wang}, X. and {Brammer}, G. and {Brada{\v{c}}}, M. and {Chen}, W. and {Kelly}, P.~L. and {Koekemoer}, A.~M. and {Trenti}, M. and {Windhorst}, R.~A.},
  title         = {{Closing in on the sources of cosmic reionization: First results from the GLASS-JWST program}},
  journal       = {\aap},
  year          = 2023,
  month         = apr,
  volume        = {672},
  eid           = {A155},
  pages         = {A155},
  doi           = {10.1051/0004-6361/202345866},
  archiveprefix = {arXiv},
  eprint        = {2301.02816},
  primaryclass  = {astro-ph.GA},
  adsurl        = {https://ui.adsabs.harvard.edu/abs/2023A&A...672A.155M}
}

@article{Trapp2023,
  author        = {{Trapp}, A.~C. and {Furlanetto}, Steven R. and {Davies}, Frederick B.},
  title         = {{Lyman {\ensuremath{\alpha}} emitters in ionized bubbles: constraining the environment and ionized fraction}},
  journal       = {\mnras},
  year          = 2023,
  month         = oct,
  volume        = {524},
  number        = {4},
  pages         = {5891-5903},
  doi           = {10.1093/mnras/stad2228},
  archiveprefix = {arXiv},
  eprint        = {2210.06504},
  primaryclass  = {astro-ph.CO},
  adsurl        = {https://ui.adsabs.harvard.edu/abs/2023MNRAS.524.5891T}
}

@article{Mason2025,
  author        = {{Mason}, Charlotte A. and {Chen}, Zuyi and {Stark}, Daniel P. and {Lu}, Ting-Yi and {Topping}, Michael and {Tang}, Mengtao},
  title         = {{Constraints on the z {\ensuremath{\sim}} 6{\ensuremath{-}}13 intergalactic medium from JWST spectroscopy of Lyman-alpha damping wings in galaxies}},
  journal       = {\aap},
  year          = 2026,
  month         = jan,
  volume        = {705},
  eid           = {A114},
  pages         = {A114},
  doi           = {10.1051/0004-6361/202553820},
  archiveprefix = {arXiv},
  eprint        = {2501.11702},
  primaryclass  = {astro-ph.GA},
  adsurl        = {https://ui.adsabs.harvard.edu/abs/2026A&A...705A.114M}
}

@article{Greene2024,
  author        = {{Greene}, Jenny E. and {Labbe}, Ivo and {Goulding}, Andy D. and {Furtak}, Lukas J. and {Chemerynska}, Iryna and {Kokorev}, Vasily and {Dayal}, Pratika and {Volonteri}, Marta and {Williams}, Christina C. and {Wang}, Bingjie and {Setton}, David J. and {Burgasser}, Adam J. and {Bezanson}, Rachel and {Atek}, Hakim and {Brammer}, Gabriel and {Cutler}, Sam E. and {Feldmann}, Robert and {Fujimoto}, Seiji and {Glazebrook}, Karl and {de Graaff}, Anna and {Khullar}, Gourav and {Leja}, Joel and {Marchesini}, Danilo and {Maseda}, Michael V. and {Matthee}, Jorryt and {Miller}, Tim B. and {Naidu}, Rohan P. and {Nanayakkara}, Themiya and {Oesch}, Pascal A. and {Pan}, Richard and {Papovich}, Casey and {Price}, Sedona H. and {van Dokkum}, Pieter and {Weaver}, John R. and {Whitaker}, Katherine E. and {Zitrin}, Adi},
  title         = {{UNCOVER Spectroscopy Confirms the Surprising Ubiquity of Active Galactic Nuclei in Red Sources at z > 5}},
  journal       = {\apj},
  year          = 2024,
  month         = mar,
  volume        = {964},
  number        = {1},
  eid           = {39},
  pages         = {39},
  doi           = {10.3847/1538-4357/ad1e5f},
  archiveprefix = {arXiv},
  eprint        = {2309.05714},
  primaryclass  = {astro-ph.GA},
  adsurl        = {https://ui.adsabs.harvard.edu/abs/2024ApJ...964...39G}
}

@article{Naidu2020,
  author        = {{Naidu}, Rohan P. and {Tacchella}, Sandro and {Mason}, Charlotte A. and {Bose}, Sownak and {Oesch}, Pascal A. and {Conroy}, Charlie},
  title         = {{Rapid Reionization by the Oligarchs: The Case for Massive, UV-bright, Star-forming Galaxies with High Escape Fractions}},
  journal       = {\apj},
  year          = 2020,
  month         = apr,
  volume        = {892},
  number        = {2},
  eid           = {109},
  pages         = {109},
  doi           = {10.3847/1538-4357/ab7cc9},
  archiveprefix = {arXiv},
  eprint        = {1907.13130},
  primaryclass  = {astro-ph.GA},
  adsurl        = {https://ui.adsabs.harvard.edu/abs/2020ApJ...892..109N}
}

@article{Bunker2024GNz11,
  author        = {{Bunker}, Andrew J. and {Saxena}, Aayush and {Cameron}, Alex J. and {Willott}, Chris J. and {Curtis-Lake}, Emma and {Jakobsen}, Peter and {Carniani}, Stefano and {Smit}, Renske and {Maiolino}, Roberto and {Witstok}, Joris and {Curti}, Mirko and {D'Eugenio}, Francesco and {Jones}, Gareth C. and {Ferruit}, Pierre and {Arribas}, Santiago and {Charlot}, Stephane and {Chevallard}, Jacopo and {Giardino}, Giovanna and {de Graaff}, Anna and {Looser}, Tobias J. and {L{\"u}tzgendorf}, Nora and {Maseda}, Michael V. and {Rawle}, Tim and {Rix}, Hans-Walter and {Del Pino}, Bruno Rodr{\'\i}guez and {Alberts}, Stacey and {Egami}, Eiichi and {Eisenstein}, Daniel J. and {Endsley}, Ryan and {Hainline}, Kevin and {Hausen}, Ryan and {Johnson}, Benjamin D. and {Rieke}, George and {Rieke}, Marcia and {Robertson}, Brant E. and {Shivaei}, Irene and {Stark}, Daniel P. and {Sun}, Fengwu and {Tacchella}, Sandro and {Tang}, Mengtao and {Williams}, Christina C. and {Willmer}, Christopher N.~A. and {Baker}, William M. and {Baum}, Stefi and {Bhatawdekar}, Rachana and {Bowler}, Rebecca and {Boyett}, Kristan and {Chen}, Zuyi and {Circosta}, Chiara and {Helton}, Jakob M. and {Ji}, Zhiyuan and {Kumari}, Nimisha and {Lyu}, Jianwei and {Nelson}, Erica and {Parlanti}, Eleonora and {Perna}, Michele and {Sandles}, Lester and {Scholtz}, Jan and {Suess}, Katherine A. and {Topping}, Michael W. and {{\"U}bler}, Hannah and {Wallace}, Imaan E.~B. and {Whitler}, Lily},
  title         = {{JADES NIRSpec Spectroscopy of GN-z11: Lyman-{\ensuremath{\alpha}} emission and possible enhanced nitrogen abundance in a z = 10.60 luminous galaxy}},
  journal       = {\aap},
  year          = 2023,
  month         = sep,
  volume        = {677},
  eid           = {A88},
  pages         = {A88},
  doi           = {10.1051/0004-6361/202346159},
  archiveprefix = {arXiv},
  eprint        = {2302.07256},
  primaryclass  = {astro-ph.GA},
  adsurl        = {https://ui.adsabs.harvard.edu/abs/2023A&A...677A..88B}
}

@article{Bosman2022,
  author        = {{Bosman}, Sarah E.~I. and {Davies}, Frederick B. and {Becker}, George D. and {Keating}, Laura C. and {Davies}, Rebecca L. and {Zhu}, Yongda and {Eilers}, Anna-Christina and {D'Odorico}, Valentina and {Bian}, Fuyan and {Bischetti}, Manuela and {Cristiani}, Stefano V. and {Fan}, Xiaohui and {Farina}, Emanuele P. and {Haehnelt}, Martin G. and {Hennawi}, Joseph F. and {Kulkarni}, Girish and {Mesinger}, Andrei and {Meyer}, Romain A. and {Onoue}, Masafusa and {Pallottini}, Andrea and {Qin}, Yuxiang and {Ryan-Weber}, Emma and {Schindler}, Jan-Torge and {Walter}, Fabian and {Wang}, Feige and {Yang}, Jinyi},
  title         = {{Hydrogen reionization ends by z = 5.3: Lyman-{\ensuremath{\alpha}} optical depth measured by the XQR-30 sample}},
  journal       = {\mnras},
  year          = 2022,
  month         = jul,
  volume        = {514},
  number        = {1},
  pages         = {55-76},
  doi           = {10.1093/mnras/stac1046},
  archiveprefix = {arXiv},
  eprint        = {2108.03699},
  primaryclass  = {astro-ph.CO},
  adsurl        = {https://ui.adsabs.harvard.edu/abs/2022MNRAS.514...55B}
}

@article{MadauDickinson2014,
  author        = {{Madau}, Piero and {Dickinson}, Mark},
  title         = {{Cosmic Star-Formation History}},
  journal       = {\araa},
  year          = 2014,
  month         = aug,
  volume        = {52},
  pages         = {415-486},
  doi           = {10.1146/annurev-astro-081811-125615},
  archiveprefix = {arXiv},
  eprint        = {1403.0007},
  primaryclass  = {astro-ph.CO},
  adsurl        = {https://ui.adsabs.harvard.edu/abs/2014ARA&A..52..415M}
}

@article{Totani2014,
  author        = {{Totani}, Tomonori and {Aoki}, Kentaro and {Hattori}, Takashi and {Kosugi}, George and {Niino}, Yuu and {Hashimoto}, Tetsuya and {Kawai}, Nobuyuki and {Ohta}, Kouji and {Sakamoto}, Takanori and {Yamada}, Toru},
  title         = {{Probing Intergalactic Neutral Hydrogen by the Lyman Alpha Red Damping Wing of Gamma-Ray Burst 130606A Afterglow Spectrum at z = 5.913}},
  journal       = {\pasj},
  year          = 2014,
  volume        = {66},
  pages         = {63},
  doi           = {10.1093/pasj/psu032},
  archiveprefix = {arXiv},
  eprint        = {1312.3934},
  primaryclass  = {astro-ph.HE}
}

@article{Fausey2025,
  author        = {{Fausey}, H. M. and {van der Horst}, A. J. and {Tanvir}, N. R. and {Wiersema}, K. and {Fynbo}, J. P. U. and {Hartmann}, D. and {de Ugarte Postigo}, A.},
  title         = {{Revisiting the Intergalactic Medium Around GRB 130606A and Constraints on the Epoch of Reionization}},
  journal       = {\apj},
  year          = 2025,
  volume        = {985},
  pages         = {28},
  doi           = {10.3847/1538-4357/adc5fc},
  archiveprefix = {arXiv},
  eprint        = {2412.09732},
  primaryclass  = {astro-ph.HE}
}

@misc{Hu2026C3PO,
  author        = {{Hu}, Weida and {Papovich}, Casey and {Hutchison}, Taylor A. and {Larson}, Rebecca L. and {Arrabal Haro}, Pablo and {Finkelstein}, Steven L. and {Cleri}, Nikko J. and {Shen}, Lu and {Yang}, Jiayang},
  title         = {{C3PO: A Ly$\alpha$ Emitting Galaxy at $z\sim9$}},
  year          = 2026,
  eid           = {arXiv:2609.03608},
  archiveprefix = {arXiv},
  eprint        = {2609.03608},
  primaryclass  = {astro-ph.GA},
  url           = {https://arxiv.org/abs/2609.03608}
}

@article{Correa2015AccretionII,
  author        = {{Correa}, Camila A. and {Wyithe}, J. Stuart B. and {Schaye}, Joop and {Duffy}, Alan R.},
  title         = {{The accretion history of dark matter haloes II: The connections with the mass power spectrum and the density profile}},
  journal       = {\mnras},
  year          = 2015,
  volume        = {450},
  pages         = {1521-1537},
  doi           = {10.1093/mnras/stv697},
  archiveprefix = {arXiv},
  eprint        = {1501.04382},
  primaryclass  = {astro-ph.CO}
}

@article{ShethTormen1999,
  author        = {{Sheth}, Ravi K. and {Tormen}, Giuseppe},
  title         = {{Large scale bias and the peak background split}},
  journal       = {\mnras},
  year          = 1999,
  volume        = {308},
  pages         = {119--126},
  doi           = {10.1046/j.1365-8711.1999.02692.x},
  archiveprefix = {arXiv},
  eprint        = {astro-ph/9901122}
}

@article{Naidu2026MoM,
  author = {{Naidu}, Rohan P. and {Oesch}, Pascal A. and {Brammer}, Gabriel and others},
  title = {{A Cosmic Miracle: A Remarkably Luminous Galaxy at $z_{\rm spec}=14.44$ Confirmed with JWST}},
  journal = {The Open Journal of Astrophysics},
  year = {2026},
  volume = {9},
  doi = {10.33232/001c.156033},
  archivePrefix = {arXiv},
  eprint = {2505.11263},
  url = {https://arxiv.org/abs/2505.11263}
}

@article{Schouws2025GSz14,
  author = {{Schouws}, Sander and {Bouwens}, Rychard J. and {Ormerod}, Katherine and others},
  title = {{Detection of [O III]$_{88\,\mu{\rm m}}$ in JADES-GS-z14-0 at $z=14.1793$}},
  journal = {\apj},
  year = {2025},
  volume = {988},
  pages = {19},
  doi = {10.3847/1538-4357/adbf1b},
  archivePrefix = {arXiv},
  eprint = {2409.20549}
}

@article{CurtisLake2023,
  author = {{Curtis-Lake}, Emma and {Carniani}, Stefano and {Cameron}, Alex and others},
  title = {{Spectroscopic confirmation of four metal-poor galaxies at $z=10.3$--$13.2$}},
  journal = {Nature Astronomy},
  year = {2023},
  volume = {7},
  pages = {622--632},
  doi = {10.1038/s41550-023-01918-w},
  archivePrefix = {arXiv},
  eprint = {2212.04568}
}

@article{Castellano2024GHZ2,
  author = {{Castellano}, Marco and {Napolitano}, Lorenzo and {Fontana}, Adriano and others},
  title = {{JWST NIRSpec Spectroscopy of the Remarkable Bright Galaxy GHZ2/GLASS-z12 at Redshift 12.34}},
  journal = {\apj},
  year = {2024},
  volume = {972},
  pages = {143},
  doi = {10.3847/1538-4357/ad5f88},
  archivePrefix = {arXiv},
  eprint = {2403.10238}
}

@article{ArrabalHaro2023,
  author = {{Arrabal Haro}, Pablo and {Dickinson}, Mark and {Finkelstein}, Steven L. and others},
  title = {{Confirmation and refutation of very luminous galaxies in the early Universe}},
  journal = {Nature},
  year = {2023},
  volume = {622},
  pages = {707--711},
  doi = {10.1038/s41586-023-06521-7},
  archivePrefix = {arXiv},
  eprint = {2303.15431}
}

@article{Tacchella2023GNz11,
  author = {{Tacchella}, Sandro and {Eisenstein}, Daniel J. and {Hainline}, Kevin and others},
  title = {{JADES Imaging of GN-z11: Revealing the Morphology and Environment of a Luminous Galaxy 430 Myr after the Big Bang}},
  journal = {\apj},
  year = {2023},
  volume = {952},
  pages = {74},
  doi = {10.3847/1538-4357/acdbc6},
  archivePrefix = {arXiv},
  eprint = {2302.07234}
}

@article{Hsiao2024NIRSpec,
  author = {{Hsiao}, Tiger Yu-Yang and {Abdurro'uf} and {Coe}, Dan and others},
  title = {{JWST NIRSpec Spectroscopy of the Triply Lensed $z=10.17$ Galaxy MACS0647--JD}},
  journal = {\apj},
  year = {2024},
  volume = {973},
  pages = {8},
  doi = {10.3847/1538-4357/ad5da8},
  archivePrefix = {arXiv},
  eprint = {2305.03042}
}

@article{Hsiao2024MIRI,
  author = {{Hsiao}, Tiger Yu-Yang and {{\'A}lvarez-M{\'a}rquez}, Javier and {Coe}, Dan and others},
  title = {{JWST MIRI Detections of H$\alpha$ and [O III] and a Direct Metallicity Measurement of the $z=10.17$ Lensed Galaxy MACS0647--JD}},
  journal = {\apj},
  year = {2024},
  volume = {973},
  pages = {81},
  doi = {10.3847/1538-4357/ad6562},
  archivePrefix = {arXiv},
  eprint = {2404.16200}
}

@article{Lu2024BSD,
  author = {{Lu}, Ting-Yi and {Mason}, Charlotte A. and {Hutter}, Anne and {Mesinger}, Andrei and {Qin}, Yuxiang and {Stark}, Daniel P. and {Endsley}, Ryan},
  title = {{The reionizing bubble size distribution around galaxies}},
  journal = {\mnras},
  year = {2024},
  volume = {528},
  pages = {4872--4890},
  doi = {10.1093/mnras/stae266},
  archivePrefix = {arXiv},
  eprint = {2304.11192},
  primaryClass = {astro-ph.GA}
}

@article{Qin2025Lya,
  author = {{Qin}, Yuxiang and {Wyithe}, J. Stuart B.},
  title = {{Reionization morphology and intrinsic velocity offsets allow transmission of Lyman-$\alpha$ emission from JADES-GS-z13-1-LA}},
  journal = {\mnras},
  year = {2025},
  volume = {538},
  pages = {L16--L23},
  doi = {10.1093/mnrasl/slaf001},
  archivePrefix = {arXiv},
  eprint = {2409.07356},
  primaryClass = {astro-ph.CO}
}

@article{Cohon2026LAE,
  author = {{Cohon}, Joshua and {Cain}, Christopher and {Windhorst}, Rogier and {D'Aloisio}, Anson and {Carleton}, Timothy and {Zhu}, Yongda},
  title = {{A long time ago in an LAE far, far away: A signpost of early reionisation or a nascent AGN at $z=13$?}},
  journal = {\pasa},
  year = {2026},
  volume = {43},
  pages = {e046},
  doi = {10.1017/pasa.2026.10166},
  archivePrefix = {arXiv},
  eprint = {2508.05739},
  primaryClass = {astro-ph.GA}
}

@article{Witstok2026OASIS,
  author = {{Witstok}, Joris and {Carniani}, Stefano and {Jakobsen}, Peter and {Bunker}, Andrew J. and {Cameron}, Alex J. and {D'Eugenio}, Francesco and {Hainline}, Kevin and {Helton}, Jakob M. and {Looser}, Tobias J. and {Rinaldi}, Pierluigi and {Robertson}, Brant and {Baker}, William M. and {Charlot}, St{\'e}phane and {Johnson}, Benjamin D. and {Jones}, Gareth C. and {Kumari}, Nimisha and {Maiolino}, Roberto and {Scholtz}, Jan and {Tacchella}, Sandro and {Willmer}, Christopher N. A. and {Willott}, Chris and {Wu}, Zihao},
  title = {{An OASIS of Lyman-$\alpha$ within a neutral intergalactic desert: reaffirmed line and blue continuum reveal efficient ionising agents at $z=13$}},
  year = {2026},
  archivePrefix = {arXiv},
  eprint = {2603.18775},
  primaryClass = {astro-ph.GA},
  doi = {10.48550/arXiv.2603.18775},
  url = {https://arxiv.org/abs/2603.18775}
}
\bibliographystyle{aasjournalv7}

\end{document}